\RequirePackage{fix-cm}
\documentclass[final]{svjour3}                                   % onecolumn (standard format)
\usepackage[11pt]{extsizes}
 \journalname{arXiv.org}

\usepackage{mathptmx}

\usepackage{latexsym}
\usepackage{graphicx}
\usepackage{natbib}
\usepackage{amsmath}
\usepackage{algorithm}
\usepackage{algorithmic}
\usepackage{todonotes}
\usepackage{enumitem}
\usepackage{lipsum}
\usepackage{rotating}
\usepackage{pdflscape}
\usepackage{longtable,array}
\usepackage{caption}
\usepackage{booktabs}
\usepackage{array,xstring}
\usepackage{rotating}
\usepackage{color}
\usepackage{epstopdf}

\usepackage[top=1.25in, bottom=1.1in, left=1.1in, right=1.1in]{geometry}

 \newcommand{\tsum}{\mathop{\textstyle \sum }}
 \newcommand{\tprod}{\mathop{\textstyle \prod}}

\titlerunning{{\small Bayesian panel quantile regression for binary outcomes with correlated random effects} }  % (should be dropped for the final version to submit to EE)
\authorrunning{Bresson, Lacroix and Rahman}                                                                              % (should be dropped for the final version to submit to EE)

\begin{document}

\title{Bayesian panel quantile regression for binary outcomes with correlated random effects:
An application on crime recidivism in Canada}
%\subtitle{An application on crime recidivism in Canada}

\author{Georges Bresson \and Guy Lacroix \and Mohammad Arshad Rahman}

\institute{Georges Bresson \at
              Department of Economics, Universit\'{e} Paris II, Paris, France.\\
             \email{georges.bresson@u-paris2.fr} \\
             \emph{Corresponding author. Department of Economics, Universit\'{e} Paris II, 12 place du Panth\'{e}on, 75231 Paris cedex 05, France (Tel.: +33 (1) 44 41 89 73).}  %  if needed
           \and
           Guy Lacroix \at
           Department of Economics, Universit\'{e} Laval, Quebec, Canada.\\
           \email{guy.lacroix@ecn.ulaval.ca}
            \and
           Mohammad Arshad Rahman \at
           Department of Economic Sciences, Indian Institute of Technology, Kanpur, India.\\
           \email{marshad@iitk.ac.in}
}

\date{January 2020}

\maketitle

\begin{abstract}
This article develops a Bayesian approach for estimating panel quantile
regression with binary outcomes in the presence of correlated random effects.
We construct a working likelihood using an asymmetric Laplace (AL) error
distribution and combine it with suitable prior distributions to obtain the
complete joint posterior distribution. For posterior inference, we propose
two Markov chain Monte Carlo (MCMC) algorithms but prefer the algorithm that
exploits the blocking procedure to produce lower autocorrelation in the MCMC
draws. We also explain how to use the MCMC draws to calculate the marginal
effects, relative risk and odds ratio. The performance of our preferred
algorithm is demonstrated in multiple simulation studies and shown to perform
extremely well. Furthermore, we implement the proposed framework to study
crime recidivism in Quebec, a Canadian Province, using a novel data from the
administrative correctional files. Our results suggest that the recently
implemented ``tough-on-crime'' policy of the Canadian government has been
largely successful in reducing the probability of repeat offenses in the
post-policy period. Besides, our results support existing findings on crime
recidivism and offer new insights at various quantiles.

%\textcolor{red}{The impact of “tough-on-crime” variables on recidivism is sizeable and relatively more important than those of other control variables.
%Interestingly, the policy appears to have impacted those in the lower quantiles much more than those located in the higher ones.}

\end{abstract}

\section{Introduction}
\label{sec:intro}

The concept of quantile regression introduced in \citet{Koenker-Basset-1978}
has captured the attention of both statisticians and econometricians,
theorists as well as applied researchers, and across school of thoughts i.e.,
Classicals (or Frequentists) and Bayesians. Quantile regression offers
several advantages over mean regression (such as robustness against outliers,
desirable equivariance properties, \emph{etc}.) and estimation methods,
particularly for cross-section data, are also well developed\footnote{Some
Classical techniques include simplex method \citep{Dantzig-1963,
Dantzig-Thapa-1997,Dantzig-Thapa-2003, Barrodale-Roberts-1973,
Koenker-dOrey-1987}, interior point algorithm
\citep{Karmarkar-1984,Mehrotra-1992} and smoothing algorithm
\citep{Madsen-Nielsen-1993, Chen-2007}. Bayesian methods using Markov chain
Monte Carlo (MCMC) algorithms for estimating quantile regression was
introduced in \citet{Yu-Moyeed-2001} and refined, amongst others, in
\citet{Kozumi-Kobayashi-2011}. A non-Markovian simulation based algorithm was
proposed in \citet{Rahman-2013}. See also \citet{Soares-Fagundes-2018} for
interval quantile regression using swarm intelligence.}. The method has been
employed in various disciplines including economics, finance, and the social
sciences \citep{KoenkerBook-2005, Davino-etal-2013}. However, the development
of quantile regression for panel data witnessed noticeable delay (more than two decades)
because of complexities in estimation. The primary challenge was that
quantiles, unlike means, are not linear operators and hence standard
differencing (or demeaning) methods are not applicable to estimation of
quantile regression. The challenges in estimation increases, if, for example,
the outcome variable is discrete (such as binary or ordinal) because
quantiles for such variables are not readily defined. Besides, modeling of
panel data brings in consideration of unobserved individual-specific
heterogeneity and the related debate on the choice of ``random-effects''
versus ``fixed-effects''. Motivated by these challenges in modeling and
estimation, this paper considers a quantile regression model for panel data
in the presence of correlated-random effects (CRE) and introduces two Markov
chain Monte Carlo (MCMC) algorithms for its estimation. The proposed
framework is applied to study crime recidivism in the Province of Quebec, Canada,
using a novel data constructed from the administrative correctional files.

The current paper touches on at least two growing econometric/statistic
literatures -- quantile regression for panel data and panel quantile
regression for discrete outcomes. In reference to the former,
\citet{Koenker-2004} was first to suggest a penalization based approach to
estimate quantile regression model with unobserved individual-specific
effects\footnote{For other development in quantile regression on panel data see, amongst
others, \citet{Lamarche-2010}, \citet{Canay-2011},
\citet{Chernozhukov-etal-2013}, \citet{Galvao-etal-2013},
\citet{Galvao-Kato-2017}, \cite{Graham-etal-2018}, and
\citet{Galvao-Poirier-2019} to mention a few.}. \citet{Geraci-Bottai-2007}
adopted the likelihood based approach of \citet{Yu-Moyeed-2001} and
constructed a working likelihood using the asymmetric Laplace (AL)
distribution. They proposed a Monte Carlo expectation-maximization (EM)
algorithm to estimate the panel quantile regression model and apply it to
study labor pain data reported in \citet{Davis-1991}. Later,
\citet{Geraci-Bottai-2014} extended the panel quantile regression model of
\citet{Geraci-Bottai-2007} to accommodate multiple individual-specific
effects and suggested strategies to reduce the computational burden of the
Monte Carlo EM algorithm. A Bayesian approach to estimate the panel quantile
regression was presented in \citet{Luo-Lian-Tian-2012}, where they propose a
Gibbs sampling algorithm by exploiting the normal-exponential mixture
representation of the AL distribution \citep{Kozumi-Kobayashi-2011}.
\citet{Wang-2012} also utilized the AL density to develop a Bayesian
estimation method for quantile regression in a parametric nonlinear
mixed-effects model.

The papers on quantile regression mentioned in the previous paragraph have
assumed that the unobserved individual-specific effects are uncorrelated with
the regressors -- also known as ``random-effects'' in the Classical
econometrics literature. In contrast, when the individual-specific effects
are assumed to be correlated with the regressors, the models have been termed
as ``fixed-effects'' model. Fixed-effects models suffer from the limitation
that it cannot estimate the coefficient for time-invariant regressors. So,
when most of the variation in a regressor is located in the individual
dimension (rather than in the time dimension), estimation of coefficients of
time varying regressors may be imprecise. Most disciplines in applied
statistics, other than econometrics, use the random-effects model
\citep{Cameron-Trivedi-2005}. However, as shown in \citet{Baltagi-2013}, most
applied work in economics have settled the choice between the two
specifications using the specification test proposed in \cite{Hausman-1978}.

Between the questionable orthogonality assumption of the random-effects model
and the limitations of the fixed-effects specification, lies the idea of
correlated random-effects (CRE). This concept is utilized in the current
paper to soften the assertion of unobserved individual heterogeneity being
uncorrelated with regressors. The CRE was introduced in \citet{Mundlak-1978},
where he models the individual-specific effects as a linear function of the
time averages of all the regressors. \citet{Hausman-Taylor-1981} proposed an
alternative specification in which some of the time-varying and
time-invariant regressors are related to the unobserved individual-specific
effects.\footnote{\citet{Baltagi-etal-2003} suggested an alternative
\textit{pretest} estimator based on the Hausman-Taylor (HT) model. This
pretest alternative considers an HT model in which some of the variables, but
not all, may be correlated with the individual effects. The pretest estimator
becomes the random-effects estimator if the standard Hausman test is not
rejected. The pretest estimator becomes the HT estimator if a second Hausman
test (based on the difference between the FE and HT estimators) does not
reject the choice of strictly exogenous regressors. Otherwise, the pretest
estimator is the FE estimator.} Later, \cite{Chamberlain1982,
Chamberlain-1984} considered a richer model and defined the
individual-specific effects as a weighted sum of the regressors. These CRE
models lead to an estimator of the coefficients of the regressors that equals
the fixed-effects estimator. The literature has numerous publications on the
Hausman tests or the CRE models in a linear or non-linear framework. We refer
the reader to \cite{Baltagi-2013}, \cite{Wooldridge-2010},
\citet{Arellano-1993}, \citet{Burda-Harding-2013}, \citet{Greene-2015} and
references therein. Most recently, \citet{Joshi-Wooldridge-2019} extended the
CRE approach to linear panel data models when instrumental variables are
needed and the panel is unbalanced.

Within the quantile regression for panel data literature,
\citet{Abrevaya-Dahl-2008} incorporated the CRE to the quantile panel
regression model and utilized it to study birth weight using a balanced panel
data from Arizona and Washington. They make certain simplifying assumptions
which allows them to estimate the model using pooled linear quantile
regression. Following the quantile regression framework of
\citet{Abrevaya-Dahl-2008}, \citet{Bache-etal-2013} considers a more
restricted specification to model birth weight using an unbalanced panel data
from Denmark. \citet{Arellano-Bonhomme-2016} introduced a class of QR
estimators for short panels, where the conditional quantile response function
of the unobserved heterogeneity is also specified as a function of
observables. The literature on Bayesian panel quantile regression with CRE is
limited to \citet{Kobayashi-Kozumi-2012}, where they develop Bayesian
quantile regression for censored dynamic panel data and proposed a Gibbs
sampling algorithm to estimate the model. The initial condition problem
arising due to the dynamic nature of the model was successfully managed using
correlated random effects. In addition, they implement the framework to study

The literature on panel quantile regression for discrete outcomes is quite
sparse and most of the work has only come recently\footnote{A body of work
related to quantile regression for discrete outcomes include, but is not
limited to, \citet{Kordas-2006}, \citet{Benoit-Poel-2010},
\citet{Alhamzawi-2016}, \citet{Omata-etal-2017}, \citet{Alhamzawi-Ali-2018}
and \citet{Rahman-Karnawat-2019}}. \citet{Alhamzawi-Ali-2018} extended the
Bayesian ordinal quantile regression introduced in \citet{Rahman-2016} to
panel data and use it to analyze treatment related changes in illness
severity using data from the National Institute of Mental Health
Schizophrenia Collaborative (NIMHSC), and previously analyzed in
\citet{Gibbons-Hedeker-1994}. \citet{Ghasemzadeh-etal-2018-METRON} proposed a
Gibbs sampling algorithm to estimate Bayesian quantile regression for ordinal
longitudinal response in the presence of non-ignorable missingness and use it
to analyze the Schizophrenia data of \citet{Gibbons-Hedeker-1994}.
\citet{Ghasemzadeh-etal-2018-Comm} developed a Bayesian quantile regression
model for bivariate longitudinal mixed ordinal and continuous responses to
study the relationship between reading ability and antisocial behavior
amongst children using the Peabody Individual Achievement Test (PIAT) data.
Most recently, \citet{Rahman-Vossmeyer-2019} considered a panel quantile
regression model with binary outcomes and develop an efficient blocked
sampling algorithm. They apply the framework to study female labor force
participation and home ownership using data from the Panel Study of Income
Dynamics (PSID).

This article contributes to the two literatures by incorporating the CRE
concept into the panel quantile regression model for binary outcomes. Our
proposed framework is more general and can accommodate the binary panel
quantile regression model of \citet{Rahman-Vossmeyer-2019} as a special case.
We present two MCMC algorithms -- a simple (non-blocked) Gibbs sampling
algorithm and another blocked Gibbs sampling algorithm that exploits the
block sampling of parameters to reduce the autocorrelation in MCMC draws. We
also explain how to calculate the marginal effects, relative risk and the
odds ratio using the MCMC draws. The performance of the blocked algorithm is
thoroughly tested in multiple simulation studies and shown to perform
extremely well. Lastly, we implement the model to study crime recidivism in
the Province of Quebec, Canada, using data from the administrative correction
files for the period 2007$-$2017. The results provide strong support for
including the CRE into the binary panel quantile regression framework. On the
applied side, we find that the recently implemented ``tough-on-crime'' policy
has been successful in reducing the probability of repeat offenses and this is
most pronounced at the lower quantiles. Besides, our results confirm existing
findings from recent studies on crime recidivism, such as, schooling
(unemployment rate) is negatively (positively) associated with crime
recidivism. Moreover, the marginal effects and relative risk show
considerable variability across the considered quantiles.

The remainder of the paper is organized as follows. Section~\ref{sec:themodel} introduces the
binary panel regression model with correlated random-effects and the two MCMC
algorithms. Section~\ref{sec:montecarlo} presents the simulation studies and discusses the
performance of the algorithm. Section~\ref{sec:ME-RR-OR} discusses how to compute the marginal
effects, relative risk and odds ratio using the MCMC draws. Section~\ref{sec:application}
implements the proposed framework to study crime recidivism in Quebec, a
Canadian Province. Section~\ref{sec:conclusion} presents concluding remarks.

%---------------------------------------------------------------------------------
\section{The Model}
\label{sec:themodel} We propose a binary quantile regression framework for
panel data where the individual-specific effects are correlated with the
covariates giving rise to correlated random effects. The resulting binary
panel quantile regression with correlated random effects (BPQRCRE) model can
be conveniently expressed in the latent variable formulation of
\citet{Albert-Chib-2001} as follows,
%-------------------------
\begin{equation}
\begin{split}
z_{it} & =  x_{it}^{\prime }\beta +\alpha _{i}+\varepsilon _{it} \hspace{0.5in} \forall \; i=1,\cdots,n, \;\; t=1,\cdots,T_{i}, \\
y_{it} & = \left\{
\begin{array}{cc}
1 & \text{ if }z_{it}>0, \\
0 & \text{otherwise,}%
\end{array}%
\right. \\
\alpha_{i} & \sim N( \overline{m}_{i}^{\prime}\zeta, \sigma _{\alpha }^{2}),
\end{split}
\label{eq:model1}
\end{equation}
%-------------------------
where $z_{it}$ is a continuous latent variable associated with the binary
outcome $y_{it}$, $x_{it}^{\prime} = (x_{it,1}, x_{it,2}, \cdots , x_{it,k})$
is a $\left( 1\times k\right) $ vector of explanatory variables including the
intercept, $\beta $ is the $\left( k\times 1\right) $ vector of common
parameters, and $\alpha _{i}$ is the individual-specific effect assumed to be
independently distributed as a normal distribution, i.e., $\alpha_{i} \sim
N\left( \overline{m}^{\prime}_{i} \zeta, \sigma _{\alpha }^{2}\right) $. Here
$\overline{m}_{i,j}=\tsum_{t=1}^{T_{i}}x_{it,j}/T_{i}$ (for $j=2,...,k$) and
$\overline{m}^{\prime}_{i} = (\overline{m}_{i,2},\cdots, \overline{m}_{i,k})$
is a $\left( 1\times (k-1)\right) $ vector of individual means of explanatory
variables excluding the intercept. The dependence of $\alpha$ on the
covariates $(x)$ yields a correlated random effects model
\citep{Mundlak-1978}. The error term $\varepsilon_{it}$, conditional on
$\alpha_{i}$, is assumed to be independently and identically distributed
(\emph{iid}) as an Asymmetric Laplace (AL) distribution i.e.,
$\varepsilon_{it}|\alpha_{i} \overset{iid}{\sim} AL\left(0,1,p\right)$, where
$p$ denotes the quantile. The AL error distribution is used to create a
working likelihood and has been utilized in previous studies on longitudinal
data models such as \citet{Luo-Lian-Tian-2012} and
\citet{Rahman-Vossmeyer-2019}.

In the proposed BPQRCRE framework, the modeling of correlated random effects
as a function of the means of the covariates is inspired from
\cite{Mundlak-1978}. Utilizing $\overline{m}^{\prime}_{i}$ as a set of
controls for unobserved heterogeneity is both intuitive and advantageous. It
is intuitive because it estimates the effect of the covariates holding the
time average fixed, and advantageous because it serves a compromise between
the questionable orthogonality assumptions of the random effects model and
the limitation of the fixed effects specification which leads to the
incidental parameters problem. The considered model reduces to the standard
uncorrelated random effects case, if we  set $\zeta=0$, i.e., assume
$\alpha_{i}$ is  independent of the covariates \citep{Rahman-Vossmeyer-2019}.
Here, we note that \cite{Chamberlain1982,Chamberlain-1984} allowed for
correlation between $\alpha _{i}$ and the covariates $x_{it}^{\prime }$
(excluding the intercept) through a more general formulation: $\alpha_{i}\sim
N\left( \tsum_{t=1}^{T_{i}}x_{it}^{\prime }\zeta _{t}, \sigma _{\alpha
}^{2}\right)$. However, this approach is more involved for an unbalanced
panel, particularly if endogeneity attrition is the reason for the panel to
be unbalanced \citep[see][]{Wooldridge-2010}. Besides, the correlated random
effects specification has a number of virtues for nonlinear panel data models
as underlined in \cite{Burda-Harding-2013} and \cite{Greene-2015}. Hence, we
prefer the approach presented in \citet{Mundlak-1978} compared to the method
in \cite{Chamberlain1980,Chamberlain1982, Chamberlain-1984}.

The BPQRCRE model as presented in equation~\eqref{eq:model1} can be directly
estimated using MCMC algorithms, but the resulting posterior will not yield
the full set of tractable conditional posteriors necessary for a Gibbs
sampler. Therefore, as done in \citet{Luo-Lian-Tian-2012} and
\citet{Rahman-Vossmeyer-2019}, we utilize the normal-exponential mixture
representation of the AL distribution to facilitate Gibbs sampling
\citep{Kozumi-Kobayashi-2011}. The mixture representation for
$\varepsilon_{it}$ can be written as follows,
%----------------------------------
\begin{equation}
\varepsilon_{it}=\theta w_{it}+\tau \sqrt{w_{it}}u_{it},  \label{eq2}
\end{equation}
%----------------------------------
where $u_{it}\sim N\left( 0,1\right) $ is mutually independent of $%
w_{it}\sim \mathcal{E}\left( 1\right) $ with $\mathcal{E}$ representing the
exponential distribution and the constants are $\theta =\frac{1-2p}{p(1-p)}$
and $\tau ^{2}=\frac{2}{p(1-p)}$. The mixture representation gives access to
the appealing properties of the normal distribution.

To implement the Bayesian approach, we stack the model across $i$. Define
$z_{i}=(z_{i1},...,z_{iT_{i}})^{\prime }$, $
y_{i}=(y_{i1},\cdots,y_{iT_{i}})^{\prime }$, $X_{i}=(x'_{i1},\cdots,x'_{iT_{i}})^{\prime}$, $w_{i}=(w_{i1},\cdots,w_{iT_{i}})^{%
\prime }$, $D_{\tau \sqrt{w_{i}}}=\tau \; \mathrm{diag}(\sqrt{w_{i1}},\cdots,\sqrt{%
w_{iT_{i}}})^{\prime }$ and $u_{i}=(u_{i1},\cdots,u_{iT_{i}})^{\prime }$. The
resulting hierarchical model can be written as,
%-----------------------------------
\begin{equation}
\begin{split}
z_{i} & =  X_{i}\beta +\iota _{T_{i}}\alpha _{i}+ w_{i} \theta + D_{\tau
\sqrt{w_{i}}} u_{i} \hspace{0.4in}  \forall \;\; i=1,...,n, \\
%----
y_{it} & = \left\{
\begin{array}{cc}
1 & \textrm{if} \; z_{it}>0, \\
0 & \text{otherwise,}
\end{array}
\right.  \hspace{1.1in} \forall \;\;  i=1,...,n,; \; t=1,...,T_{i}, \\
%-----
\alpha _{i} & \sim N\left( \overline{m}_{i}^{\prime} \zeta,
\sigma_{\alpha }^{2}\right) \hspace{0.45in}
w_{it} \sim \mathcal{E}(1),  \hspace{0.75in}  u_{it} \sim N\left( 0,1\right),\\
%-----
\beta & \sim N_{k} \left( \beta _{0}, B_{0}\right)  \hspace{0.45in}
\sigma _{\alpha }^{2} \sim IG\left( \frac{c_{1}}{2}, \frac{d_{1}}{2}\right), \hspace{0.3in} \zeta \sim N_{k-1}\left( \zeta_{0}, C_{0}\right),
\end{split}
\label{eq:model2}
\end{equation}
%----------------------------------
where $\iota _{T_{i}}$ is a $\left( T_{i}\times 1\right) $ vector of ones and
the last line in equation~\eqref{eq:model2} presents the prior distribution
on the parameters. The notation $N_{k}(\cdot)$ denotes a multivariate normal
distribution of dimension $k$ and $IG(\cdot)$ denotes an inverse-gamma
distribution. We note that the form of the prior distribution on $\beta $
holds a penalty interpretation on the quantile loss function
\citep{Koenker-2004}. A normal prior on $\beta$ implies an $\ell_{2}$ penalty
and has been used in \cite{Geraci-Bottai-2007}, \citet{yuan2010},
\citet{Luo-Lian-Tian-2012} and \cite{Rahman-Vossmeyer-2019}.

By Bayes' theorem, we express the \textquotedblleft complete joint
posterior\textquotedblright\ density as proportional to the product of
complete likelihood function and the prior distributions as follows,
%----------------------------------
\begin{equation}
\begin{split}
\pi ( \beta ,\alpha ,z,w,\zeta ,\sigma_{\alpha }^{2}\mid y)
& \propto  \bigg\{
\tprod \limits_{i=1}^{n} f( y_{i}\mid z_{i},\beta ,\alpha
_{i},w_{i},\zeta ,\sigma _{\alpha }^{2})
\pi ( z_{i}\mid \beta ,\alpha _{i},w_{i},\zeta ,\sigma _{\alpha
}^{2} )  \\
%----
& \qquad \times \pi ( w_{i} ) \pi ( \alpha _{i} )
\bigg\} \pi( \beta ) \pi( \zeta ) \pi (\sigma_{\alpha }^{2})  \\
%-----
& \propto  \bigg\{
\tprod\limits_{i=1}^{n} \bigg [ \tprod\limits_{t=1}^{T_{i}} f ( y_{it}\mid
z_{it} ) \bigg]
\pi ( z_{i}\mid \beta ,\alpha_{i}, w_{i},\zeta,\sigma_{\alpha}^{2} )
 \pi( w_{i}) \pi ( \alpha _{i})
\bigg\} \\
%-----
& \qquad \times \pi ( \beta ) \pi( \zeta) \pi( \sigma_{\alpha }^{2}),
\end{split}
\label{eq4}
\end{equation}
%----------------------------------
where the first line assumes independence between prior distributions and
second line follows from the fact that given $z_{it}$, the observed $y_{it}$
is independent of all parameters because the second line of equation~\eqref{eq:model2}
determines $y_{it}$ given $z_{it}$ with probability $1$. Substituting the
distribution of the variables associated with the likelihood and the prior
distributions in equation~\eqref{eq4} yields the following expression,%
%---------------------------------
\begin{equation}
\begin{split}
& \pi( \beta ,\alpha ,z,w,\zeta ,\sigma_{\alpha }^{2} \mid y)
\propto \bigg\{
\tprod\limits_{i=1}^{n}\tprod\limits_{t=1}^{T_{i}} \Big[ I(
z_{it}>0) I( y_{it}=1) + I( z_{it} \leq 0) I(y_{it}=0) \Big] \bigg\} \\
%-----
& \quad \times  \exp
\bigg[ -\frac{1}{2} \tsum\limits_{i=1}^{n} \Big\{ ( z_{i}-X_{i} \beta
-\iota _{T_{i}}\alpha _{i} - w_{i} \theta )^{\prime} D_{\tau \sqrt{w_{i}}}^{-2}
( z_{i}-X_{i}\beta -\iota _{T_{i}}\alpha _{i}- w_{i} \theta )
\Big\} \bigg] \\
%-----
& \quad \times \exp \bigg(
-\tsum\limits_{i=1}^{n}\tsum\limits_{t=1}^{T_{i}}w_{it}\bigg) \big( 2\pi
\sigma _{\alpha }^{2}\big)^{-\frac{n}{2}} \exp \bigg[ -\frac{1}{2\sigma
_{\alpha }^{2}}\tsum\limits_{i=1}^{n} ( \alpha _{i}-\overline{m}%
_{i}^{\prime} \zeta)^{\prime} ( \alpha _{i}-\overline{m}_{i}^{\prime}
\zeta) \bigg]  \\
%-----
& \quad \times \left( 2\pi \right) ^{-\frac{k}{2}}\left\vert B_{0}\right\vert ^{-%
\frac{1}{2}}\exp \left[ -\frac{1}{2}\left( \beta -\beta _{0}\right) ^{\prime
}B_{0}^{-1}\left( \beta -\beta _{0}\right) \right]
\left( 2\pi \right) ^{-\frac{k-1}{2}}\left\vert C_{0}\right\vert^{-%
\frac{1}{2}} \\
%-----
& \quad \times \exp \left[ -\frac{1}{2}\left( \zeta -\zeta_{0}\right)
^{\prime }C_{0}^{-1}\left( \zeta - \zeta_{0}\right) \right] \times
\left( \sigma _{\alpha }^{2}\right) ^{-\left( \frac{c_{1}}{2}+1\right) }\exp %
\left[ -\frac{d_{1}}{2\sigma _{\alpha }^{2}}\right].
\end{split}
\label{eq5}
\end{equation}%
%---------------------------------
The complete joint posterior density in equation~\eqref{eq5} does not have a
tractable form, and thus simulation techniques are necessary for estimation.
Similar to \citet{Rahman-Vossmeyer-2019}, we adopt a Bayesian approach due to
the following two reasons.. First, the likelihood function of a discrete
panel data model is analytically intractable which makes optimization
difficult using standard hill-climbing techniques. Second, numerical
simulation methods for discrete panel data models are often slow and
difficult to implement as noted in \citet{Burda-Harding-2013} and others. The
complete joint posterior distribution (equation~\ref{eq5}) readily yields a
full set of conditional distributions (outlined below) which can be readily
employed to estimate the model using Gibbs sampling.

We can derive the conditional posteriors of the parameters and latent
variables from the joint posterior density~\eqref{eq5} by a straightforward
extension of the non-blocked sampling method presented in
\citet{Rahman-Vossmeyer-2019}. This is presented in
Algorithm~\ref{algorithm:one}, and the derivations of the conditional
posterior densities can be found in the supplementary material. The
parameters $\beta$ are sampled from an updated multivariate normal
distribution. Similarly, the parameters $\alpha_i$ are sampled  from an
updated multivariate normal distribution. The latent weights $w_{it}$ are
sampled element wise from a generalized inverse Gaussian ($GIG$) distribution
\citep{Devroye-2014}. The variance $\sigma^{2}_{\alpha}$ is sampled  from an
updated inverse-gamma ($IG$) distribution. The parameters $\zeta$ are sampled
from an updated multivariate normal distribution. Last, the latent variable
$z_{it}$ is sampled element wise from an univariate truncated normal ($TN$)
distribution. Note that while drawing each of the parameters or latent
variables, we hold the remaining quantities fixed as presented in
Algorithm~\ref{algorithm:one}.

%------------------------------------------------------------------------------
\begin{algorithm}[t!]
\caption{Non-blocked sampling in the BPQRCRE model}
\begin{algorithmic}
\label{algorithm:one}\item[]
%------------------------
\begin{enumerate}
\item Sample $\beta \mid \alpha, z, w  \sim N_{k}\left( \widetilde{\beta}, \widetilde{B}\right)$ where,\newline
    $\widetilde{B}^{-1}= \left(  \tsum\limits_{i=1}^{n} X^{\prime}_{i} D_{\tau \sqrt{w_{i}}}^{-2}  X_{i}  + B^{-1}_{0} \right)$, \; and \; $\widetilde{\beta}= \widetilde{B} \left(  \tsum\limits_{i=1}^{n} X^{\prime}_{i} D_{\tau \sqrt{w_{i}}}^{-2}
    \left( z_{i} - \iota _{T_{i}}\alpha _{i} - w_{i}    \theta  \right)  + B^{-1}_{0} \beta_0 \right)$.
%------------------------
\item Sample $\alpha_i \mid \beta, z, w, \sigma _{\alpha }^{2}, \zeta   \sim N\left( \widetilde{a}, \widetilde{A}\right)$ for $i=1,\cdots,n$, where,\\
   $\widetilde{A}^{-1} = \left( \iota^{\prime}_{T_{i}} D_{\tau \sqrt{w_{i}}}^{-2} \iota _{T_{i}} +  \sigma_{\alpha }^{-2}  \right)$, \; and \; $\widetilde{a}= \widetilde{A} \left(  \iota^{\prime}_{T_{i}} D_{\tau \sqrt{w_{i}}}^{-2}
    \left( z_{i}-X_{i}\beta - w_{i}    \theta  \right)  +   \sigma _{\alpha }^{-2}  \overline{m}_{i}^{\prime }\zeta  \right)$ .
%------------------------
\item Sample $w_{it} \mid \beta, \alpha_i , z_{it}   \sim GIG\left( \frac{1}{2}, \widetilde{\lambda}_{it}, \widetilde{\eta} \right)$ for $i=1,\cdots,n$ and $t=1,\cdots,T_i$, where,\\
     $\widetilde{\lambda}_{it} = \left(  \frac{   z_{it}-x^{\prime}_{it} \beta  -\alpha_i}{   \tau } \right)^2$, \; and \; $\widetilde{\eta}= \left( \frac{\theta^2}{\tau^2} + 2 \right)$.
%------------------------
\item Sample $\sigma_{\alpha }^{2} \mid \alpha, \zeta   \sim IG\left( \frac{ \widetilde{c}_1}{2}, \frac{ \widetilde{d}_1}{2} \right)$ where,\\
$\widetilde{c}_1 = (n + c_1 )$, \; and \; $\widetilde{d}_1= d_1 + \tsum\limits_{i=1}^{n}  \left( \alpha _{i}-\overline{m}_{i}^{\prime }\zeta \right)^{\prime }\left( \alpha _{i}-\overline{m}_{i}^{\prime }\zeta \right) $.
%------------------------
\item Sample $\zeta \mid \alpha, \sigma_{\alpha }^{2} \sim N_{k-1}\left( \widetilde{\zeta}, \widetilde{\Sigma}_{\zeta}\right)$ where,\\
        $\widetilde{\Sigma}^{-1}_{\zeta}  = \left(   \sigma_{\alpha }^{-2}  \tsum\limits_{i=1}^{n} \overline{m}_{i} \overline{m}_{i}^{\prime } +  C^{-1}_{0} \right)$, \;
         and \; $\widetilde{\zeta}=   \widetilde{\Sigma}_{\zeta} \left(   \sigma _{\alpha }^{-2}  \tsum\limits_{i=1}^{n} \overline{m}_{i} \alpha^{\prime}_{i} +  C^{-1}_{0} \zeta_{0}\right)$.
%------------------------
 \item Sample the latent variable $z \mid \beta, \alpha, w$ for all values of $i=1,\cdots,n$ and $t=1,\cdots,T_i$ from an univariate truncated normal (TN) distribution as follows,  \vspace{5px} \\
%------------------------
%\begin{equation*}
%\begin{split}
$z_{it} \mid  \beta, \alpha, w  \sim  \left\{
\begin{array}{cc}
TN_{\left(- \infty , 0 \right] } \left( x^{\prime}_{it} \beta  + \alpha_i  + w_{it} \theta , \tau^2 w_{it} \right) & \text{ if } y_{it}=0, \vspace{2px}\\
TN_{\left( 0, \infty \right) } \left( x^{\prime}_{it} \beta  + \alpha_i  + w_{it} \theta , \tau^2 w_{it} \right) & \text{ if }y_{it}=1.
\end{array}
\right.$
%\end{split}
%\end{equation*}
%------------------------
\end{enumerate}
\end{algorithmic}
\end{algorithm}
%------------------------------------------------------------------------------

The MCMC procedure presented in Algorithm~\ref{algorithm:one} exhibits the
conditional posterior distributions for the parameters and latent variables
necessary for a Gibbs sampler. While this Gibbs sampler is straightforward,
there is potential for poor mixing of the MCMC draws due to correlation
between ($\beta $, $\alpha_{i}$) and ($z_{i}$, $\alpha _{i}$). This
correlation arises because the variables corresponding to the parameters in
$\alpha _{i}$ are often a subset of those in $x_{it}^{\prime }$. Thus
conditioning these items on one another leads to high autocorrelation in MCMC
draws as demonstrated in \citet{Chib-Carlin-1999} and noted in
\citet{Rahman-Vossmeyer-2019}.

%------------------------------------------------------------------------------
\begin{algorithm}[t!]
\caption{Blocked sampling in the BPQRCRE model}
\begin{algorithmic}
\label{algorithm:two}\item[]

\begin{enumerate}
\item Sample $(\beta, z_i)$ marginally of $\alpha$ in one block as follows.
\begin{enumerate}
  \item Let $\Omega_i = \sigma_{\alpha}^{2} J_{T_i} + D_{\tau \sqrt{w_{i}}}^{2}$ with
  $J_{T_i} = \iota_{T_{i}}  \iota^{\prime}_{T_{i}}$. Sample $\beta \mid z, w,\sigma_{\alpha }^{2}, \zeta \sim N_{k} \left( \widetilde{\beta}, \widetilde{B}\right)$ where, \\
  $\widetilde{B}^{-1}= \left( \tsum\limits_{i=1}^{n} X^{\prime}_{i} \Omega^{-1}_{i}  X_{i}  + B^{-1}_{0} \right)$, \; and \; $\widetilde{\beta}= \widetilde{B} \left( \tsum\limits_{i=1}^{n} X^{\prime}_{i} \Omega^{-1}_{i} \left( z_{i} - \iota _{T_{i}}  \overline{x}_{i}^{\prime } \zeta -  w_{i}    \theta  \right)  + B^{-1}_{0} \beta_0 \right)$.
%------------------------
\item Sample the vector $z_{i} \mid  \beta, w_i, \sigma _{\alpha }^{2}, \zeta \sim TMVN_{B_i} \left( X_i \beta +  \iota_{T_{i}}    \overline{m}_{i}^{\prime } \zeta + w_{i} \theta , \Omega_i   \right)$ for all $i=1,...,n$,
       where $B_i=\left( B_{i1} \times  B_{i2} \times ... \times B_{iT_i} \right)$ and $B_{it}$ is the interval $\left(0, \infty \right)$ if $y_{it}=1$ and the interval $\left(-\infty, 0 \right]$ if $y_{it}=0$. This is achieved by sampling $z_i$ at the $j$-th pass of the MCMC iteration using a series of conditional posterior distributions as follows:\vspace{5px} \\
       $z^{j}_{it} \mid  z^{j}_{i1}, ..., z^{j}_{i(t-1)},z^{j}_{i(t+1)}, ..., z^{j}_{iT_i} \sim TN_{B_{it}} \left(  \mu_{t \mid -t} ,  \Sigma_{t \mid -t} \right)$, for $t=1,...,T_i$, \vspace{5px}\\
       where $TN$ denotes a truncated normal distribution. The terms $ \mu_{t \mid -t}$ and $ \Sigma_{t \mid -t}$ are the conditional mean and variance, and are defined as,\\
       $
       \begin{array}{lll}
          \mu_{t \mid -t} & = & x^{\prime}_{it} \beta + \overline{m}_{i}^{\prime } \zeta + w_{it}\theta + \Sigma_{t, -t} \Sigma^{-1}_{-t , -t} \left( z^{j}_{i,-t}  - \left(  X_i \beta +  \iota _{T_{i}}    \overline{x}_{i}^{\prime } \zeta + w_{i} \theta  \right)_{-t}    \right), \\
           \Sigma_{t \mid -t} & = &  \Sigma_{t, t} -  \Sigma_{t, -t} \Sigma^{-1}_{-t , -t} \Sigma_{-t, t},
        \end{array}
       $\\
       where $z^{j}_{i,-t}=\left( z^{j}_{i1}, ..., z^{j}_{i(t-1)},  z^{j-1}_{i(t+1)}, ..., z^{j-1}_{iT_i} \right)^{\prime}$, $ \left(  X_i \beta +  \iota_{T_{i}}  \overline{m}_{i}^{\prime } \zeta + w_{i} \theta  \right)_{-t}$ is a column vector with $t$-th element removed, $ \Sigma_{t, t}$ denotes the $(t,t)$-th element of $\Omega_i$, $\Sigma_{t, -t}$ denotes the $t$-th row of  $\Omega_i$ with element in the $t$-th column removed and $\Sigma_{-t, -t}$ is the $\Omega_i$ matrix with $t$-th row and $t$-th column removed.
\end{enumerate}
%------------------------
\item Sample $\alpha_i \mid \beta, z, w, \sigma_{\alpha }^{2}, \zeta  \sim N\left( \widetilde{a}, \widetilde{A}\right)$ for $i=1,...,n$, where,\\
   $\widetilde{A}^{-1} = \left(\iota^{\prime}_{T_{i}} D_{\tau \sqrt{w_{i}}}^{-2} \iota_{T_{i}} +
   \sigma_{\alpha }^{-2}  \right)$, \; and \; $\widetilde{a}= \widetilde{A} \left(  \iota^{\prime} _{T_{i}} D_{\tau \sqrt{w_{i}}}^{-2}  \left( z_{i}-X_{i}\beta - w_{i} \theta  \right)
   + \sigma_{\alpha }^{-2}  \overline{m}_{i}^{\prime }\zeta  \right)$.
%------------------------
\item Sample $w_{it} \mid \beta, \alpha_i , z_{it}   \sim GIG\left( \frac{1}{2}, \widetilde{\lambda}_{it}, \widetilde{\eta} \right)$ for $i=1,...,n$, and $t=1,...,T_i$, where,\\
         $\widetilde{\lambda}_{it} = \left(  \frac{   z_{it}-x^{\prime}_{it} \beta  -\alpha_i}{ \tau} \right)^2$, \; and \; $\widetilde{\eta}= \left( \frac{\theta^2}{\tau^2} + 2 \right)$.
%------------------------
\item Sample $\sigma_{\alpha }^{2} \mid \alpha, \zeta \sim IG\left( \frac{ \widetilde{c}_1}{2}, \frac{ \widetilde{d}_1}{2} \right)$ where,\\
$\widetilde{c}_1 = (n + c_1 )$, \; and  \; $\widetilde{d}_1= d_1 + \tsum\limits_{i=1}^{n}  \left( \alpha _{i}-\overline{m}_{i}^{\prime }\zeta \right)^{\prime }\left( \alpha _{i}-\overline{m}_{i}^{\prime }\zeta \right) $.
%------------------------
\item Sample $\zeta \mid \alpha, \sigma_{\alpha }^{2} \sim N_{k-1}\left( \widetilde{\zeta}, \widetilde{\Sigma}_{\zeta}\right)$ where,\\
$\widetilde{\Sigma}^{-1}_{\zeta}  = \left(   \sigma _{\alpha }^{-2}  \tsum\limits_{i=1}^{n} \overline{m}_{i} \overline{m}_{i}^{\prime } +  C^{-1}_{0} \right)$, \;
 and \; $\widetilde{\zeta}=   \widetilde{\Sigma}_{\zeta} \left(   \sigma _{\alpha }^{-2} \tsum\limits_{i=1}^{n} \overline{m}_{i} \alpha^{\prime}_{i} +  C^{-1}_{0} \zeta_{0}\right)$.
%------------------------
\end{enumerate}
\end{algorithmic}
\end{algorithm}
%------------------------------------------------------------------------------

To avoid the high autocorrelation in MCMC draws, we present an alternative
algorithm that jointly samples ($\beta $, $z$) in one block within the Gibbs
sampler \citep[see][for more on the blocking
procedure]{Rahman-Vossmeyer-2019}. The details of our blocked sampler are
described in Algorithm~\ref{algorithm:two}, and the derivations of the
conditional posterior densities are presented in the supplementary file.
Specifically, $\beta$ is sampled marginally of $\alpha_i$ from a multivariate
normal distribution. Then the latent variable $z_i$ is sampled marginally of
$\alpha_i$ from a truncated multivariate normal distribution denoted by
$TMVN_{B_i} $, where $B_i$ is the truncation region given by $B_i=\left(
B_{i1} \times  B_{i2} \times ... \times B_{iT_i} \right)$ such that $B_{it}$
is the interval $\left(0, \infty \right)$ if $y_{it}=1$ and the interval
$\left(-\infty, 0 \right]$ if $y_{it}=0$. To draw from a truncated
multivariate normal distribution, we utilize the method proposed in
\cite{geweke1991,geweke2005}; as done in \cite{Rahman-Vossmeyer-2019}. This
involves drawing from a series of conditional posteriors which are univariate
truncated normal distributions. The parameter $\alpha_i$ is sampled
conditional on $(\beta, z, w, \sigma^{2}_{\alpha},\zeta)$ from an updated
multivariate normal distribution. The latent weights $w_{it}$ are sampled
element wise from a generalized inverse Gaussian ($GIG$) distribution
\citep{Devroye-2014}. The variance $\sigma^{2}_{\alpha}$ is sampled  from an
updated inverse-gamma ($IG$) distribution. Lastly, the parameters $\zeta$ are
sampled from an updated multivariate normal distribution. Once again, while
sampling each quantity of interest, we hold the remaining parameters or
latent variables fixed as exhibited in Algorithm~\ref{algorithm:two}.

%-------------------------------------------------------------------------------------------------
%-------------------------------------------------------------------------------------------------
\section{A Monte Carlo simulation study}\label{sec:montecarlo}

In this section, we present two simulation studies to demonstrate the
performance of the blocked algorithm for the BPQRCRE model. The simulation
data are generated from the following model,
%-----------------------------------
\begin{equation}
\begin{split}
z_{it} & =  x^{\prime}_{it}\beta + \alpha _{i}+ \varepsilon_{it}, \qquad \forall \; i=1,\cdots,n,  \; \textrm{and} \; t=1,\cdots,T_i, \\
\alpha _{i} & = \overline{m}_{i}^{\prime }\zeta + \xi_i,  \hspace{0.65in} \xi_{i}\sim N\left(0,\sigma_{\alpha }^{2}\right).
\end{split}
\label{eq6}
\end{equation}
%-----------------------------------
where $x^{\prime}_{it}=\left[ 1, \, x_{it,2}, \, x_{it,3}, \, x_{it,4}  \right]$, $\overline{m}^{\prime}_{i} =\left[ \overline{m}_{i,3}, \, \overline{m}_{i,4} \right]$,  $\overline{m}_{i,j}=\sum_{t=1}^{T_{i}}x_{it,j}/T_{i}$, $j=3,4$, $\beta=\left(\beta_{1}, \, \beta_{2}, \, \beta_{3}, \, \beta_{4} \right)^{\prime}=\left(0.5, \, 1, \, 0.6, \, -0.8\right)^{\prime}$, $\zeta=\left(\zeta_3, \, \zeta_4 \right)^{\prime} =\left(-1, 1\right)^{\prime}$. The covariates are generated as $x_{it,2} \sim U(-2,2)$, $x_{it,3} \sim U(-2,2)$, $x_{it,4} \sim U(-2,2)$, where $U$ denotes a uniform distribution, and $\sigma_{\alpha }^{2}=1$. Our first sample is unbalanced with $n=1,000$ and $T_i \sim U(5,15)$, leading to $T=\sum^{n}_{i=1} T_{i} = 9,989$ observations. In a second exercise, we increase the number of individuals $n = 2,000$ leading to $T=19,985$ observations. The error term is generated from a standard AL distribution, i.e., $\varepsilon_{it} \sim AL(0, 1, p)$ for $i=1,\cdots,n$, and $t=1,\cdots,T_i$ at three different quantiles $p$ = $0.25$, $0.5$, $0.75$.

%----------------- begin tables -------------------------

%---------------------------------------------------------------------------------------------------------
\begin{table}[b!]
   \centering \setlength{\extrarowheight}{1.5pt}
   \setlength{\tabcolsep}{5pt}
{\footnotesize
\begin{tabular}{l r rrr r rrrr rrr}
\hline
    && \multicolumn{3}{c}{25th Quantile} && \multicolumn{3}{c}{50th Quantile} && \multicolumn{3}{c}{75th Quantile} \\
  \cline{3-5} \cline{7-9} \cline{11-13}
   && Lag $1$ &  Lag $5$ &  Lag $10$ &&  Lag $1$ &  Lag $5$ &  Lag $10$  &&  Lag $1$ &  Lag $5$ &  Lag $10$ \\
\hline
\multicolumn{13}{c}{n=1000} \\
\hline
$\beta_{1}$   &&  0.1351 & 0.0338 & $-0.0258$  && 0.0544 & $-0.0079$ & $-0.0382$  && $-0.0417$ & $-0.0652$ & 0.0165 \\
$\beta_{2}$   && 0.3066 & 0.0369 & 0.0161  && 0.2385 & 0.0099 & $-0.0218$  &&  0.2688 & $-0.0567$ & 0.0253 \\
$\beta_{3}$   && 0.2828 & 0.0730 & $-0.0003$  && 0.1745 & 0.0012 & $-0.0228$  && 0.1784 & $-0.0215$ & $-0.0125$ \\
$\beta_{4}$   && 0.3372 & 0.0783 & 0.0179  && 0.2421 & 0.0037 & 0.0348  && 0.1871 & $-0.0254$ & $-0.0617$ \\
$\zeta_{3}$   && 0.0653 & 0.0160 & $-0.0314$  && 0.0389 & $-0.0080$ & 0.0034 && 0.0669 & $-0.0388$ & $-0.0338$ \\
$\zeta_{4}$   && 0.1438 & 0.0319 & $-0.0252$  &&  0.0649 & $-0.0217$ & $-0.0721$  && 0.0793 & $-0.0317$ & 0.0362 \\
$\sigma_{\alpha }^{2}$   && 0.4439 & 0.0658 & 0.0274   && 0.3115 & $-0.0004$ & $-0.0181$  && 0.3122 & 0.0151 & $-0.0050$ \\
\hline
\multicolumn{13}{c}{n=2000} \\
\hline
$\beta_{1}$   &&  0.1353 & 0.0200 & $-0.0296$  && 0.0176 & 0.0207 & 0.0154  && 0.0200 & 0.0096 & 0.0134 \\
$\beta_{2}$   && 0.3092 & 0.0035 & 0.0189 && 0.3022 & $-0.0151$ & $-0.0640$  &&  0.2539 & $-0.0229$ & $-0.0079$  \\
$\beta_{3}$   &&  0.1679 & 0.0655 & 0.0404  && 0.2051 & $-0.0201$ & $-0.0142$ &&  0.2171 & 0.0367 & 0.0325 \\
$\beta_{4}$   &&  0.2648 & 0.0359 & 0.0222  && 0.2634 & 0.0415 & $-0.0073$  && 0.1816 & 0.0262 & 0.0575 \\
$\zeta_{3}$   &&  0.0328 & $-0.0098$ & 0.0132 && 0.0762 & $-0.0567$ & 0.0017 && 0.0553 & 0.0340 & $-0.0261$ \\
$\zeta_{4}$   && 0.0782 & $-0.0415$ & $-0.0178$  &&   0.0117 & 0.0179 & 0.0137  && 0.0314 & 0.0198 & $-0.0022$ \\
$\sigma_{\alpha }^{2}$   &&  0.4381 & 0.0423 & 0.0227  && 0.3139 & $-0.0189$ & 0.0215  && 0.4017 & $-0.0072$ & 0.0621 \\
\hline
\end{tabular}
%\label{table1}
%\captionsetup{labelformat=empty, justification=justified}
%	\parbox{0.95\linewidth}{\caption{\footnotesize{Autocorrelation in MCMC draws at Lag $1$, Lag $5$ and Lag $10$ for $n=1,000$ individuals (upper panel) and $n=2,000$ individuals (lower panel).}}}
		\parbox{\linewidth}{\caption{\label{table1}\footnotesize{Autocorrelation in MCMC draws at Lag $1$, Lag $5$ and Lag $10$ for $n=1,000$ individuals (upper panel) and $n=2,000$ individuals (lower panel).}}}

}
\end{table}
%----------------------------------------------------------------------------------

%--------------------------------------------------------------------------------
\begin{figure}[h!]
\centering
\makebox[\textwidth]{
\includegraphics[width=6.75in, height = 8.0in, trim = {0 1.4cm 0 1cm}, clip]{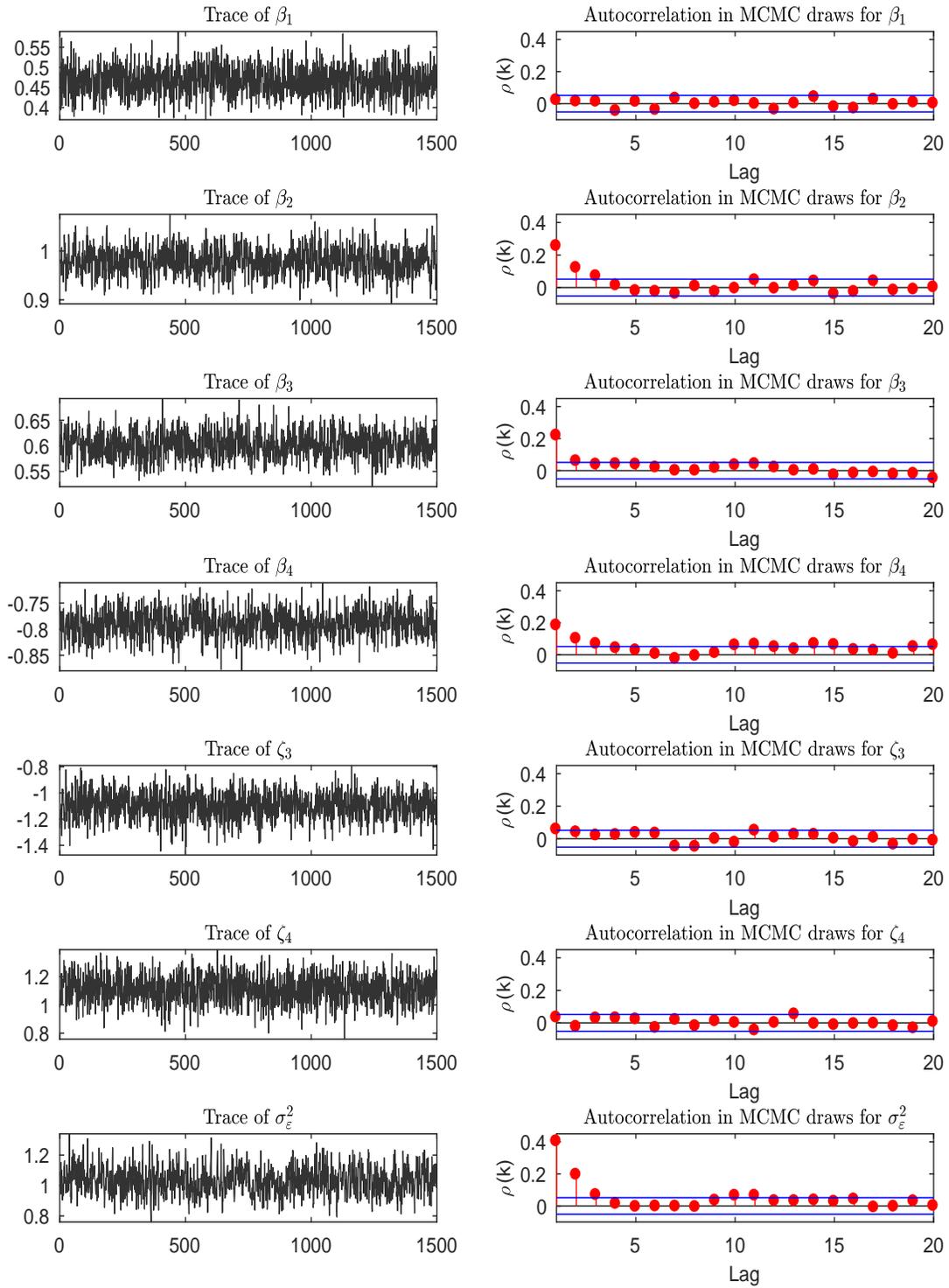}
}
\parbox{\linewidth}{\caption{\label{fig:figure1}\footnotesize{Trace plots and autocorrelation plots of the parameters for the $75$th quantile and $n=2,000$ individuals.}}}

\end{figure}
%--------------------------------------------------------------------------------

The binary outcome variable $y$ is constructed from the continuous variable
$z$, by assigning $y_{it}=1$ whenever $z_{it}>0$ and $y_{it}=0$ whenever
$z_{it} \le 0$ for all of $i=1,\cdots,n$ and $t=1,\cdots,T_{i}$. We note that
the binary response values of $0$s and $1$s are different at each quantile,
because the error values generated from an AL distribution are different for
each quantile. In the first simulation exercise with $n=1,000$, the number of
observations corresponding to $0$s and $1$s for the $25$th, $50$th and $75$th
quantiles are $(2283, 7706)$, $(4217, 5772)$ and $(6442, 3547)$,
respectively. In the second simulation exercise with $n=2,000$, the number of
observations corresponding to $0$s and $1$s for the $25$th, $50$th and $75$th
quantiles are $(4640, 15345)$, $(8691, 11294)$ and $(13234, 6751)$,
respectively. To complete the Bayesian setup for estimation, we use the
following independent prior distributions: $\beta \sim N_{k}\left( 0_{k} ,
10^3 I_{k} \right)$, $\zeta \sim N_{k-1}\left( 0_{k-1} , 10^3 I_{k-2}
\right)$, $\sigma_{\alpha }^{2} \sim IG \left( 10/2 , 9/2 \right)$. For each
exercise, we generate $16,000$ MCMC samples where the first $1,000$ values
are discarded as burn-ins. The posterior estimates are reported based on the
remaining $15,000$ MCMC iterations with a thinning factor of $10$. The mixing
of the MCMC chain is extremely good as illustrated in
Figure~\ref{fig:figure1}, which reports the trace and autocorrelation plots
of the parameters from the second simulation exercise at the 75th quantile.
The figure shows that, as desired, the chains mix well and the
autocorrelation of the MCMC draws are close to zero. The plots from the first
simulation exercise and the remaining quantiles in the second simulation
exercise are extremely similar and not presented to avoid repetition and keep
the paper within reasonable length. To supplement the plots in Figure~1,
Table \ref{table1} presents the autocorrelation in MCMC draws at lag $1$, lag
$5$, and lag $10$ confirming the good mixing across simulation exercises and
at all quantiles.

The results from the two simulation exercises are presented in
Table~\ref{table2}. Specifically, the table reports the true values of the
parameters used to generate the data, along with the posterior mean, standard
deviation and inefficiency factor \citep[calculated using the batch-means
method discussed in][]{Greenberg-2012} of the MCMC draws. In general, the
results show that the posterior means for $(\beta, \zeta)$ are near to their
respective true values, $\beta=\left(0.5,1, 0.6, -0.8\right)^{\prime}$ and
$\zeta=\left(-1,1\right)^{\prime}$ across all considered quantiles. The
posterior standard deviations for all the parameters are small and all the
coefficients are statistically different from zero. So, the proposed MCMC
algorithm is successful in correctly estimating all the model parameters
across all quantiles. This is especially important because the number of $0$s
and $1$s were different for each quantile. Moreover, the inefficiency factor
for all the parameters is close to $1$, suggesting a good sampling
performance and a nice mixing of the Markov chain. Comparing the results from
the first and second simulation exercise, we see that when the sample size is
increased from ($n=1,000$, $T=9,989$) to ($n=2,000$, $T=19,985$), the results
improve and the posterior means of the coefficients are closer to their true
values. In particular, some small observed biases for $\beta_1$, $\zeta_3$,
and $\zeta_4$ at the 25th quantile are reduced to a large extent. To
summarize, the proposed algorithm for estimating BQQRCRE model does well in
both the simulations, but the advantages of having a larger data is clearly
evident in the posterior results.

%----------------------------------------------------------------------------------
\begin{table}[t!]
   \centering \setlength{\extrarowheight}{1.5pt}
   \setlength{\tabcolsep}{5pt}
{\footnotesize
\begin{tabular}{lr rrr rrr rrr rrr}
\hline
 &  && \multicolumn{3}{c}{25th Quantile} && \multicolumn{3}{c}{50th Quantile} && \multicolumn{3}{c}{75th Quantile} \\
  \cline{4-6} \cline{8-10} \cline{12-14}
& \textsc{true} && \textsc{mean} & \textsc{std} & \textsc{if}
&& \textsc{mean} & \textsc{std} & \textsc{if}
&& \textsc{mean} & \textsc{std} & \textsc{if}\\
\hline
& \multicolumn{13}{c}{n=1000} \\
\hline
$\beta_{1}$  & 0.5 && 0.7155 & 0.0582 &  1.2290  && 0.5799 &  0.0480 & 1.0544  && 0.5319  & 0.0523 &  0.9583  \\
$\beta_{2}$  & 1.0 && 1.0093 & 0.0437 & 1.7885   && 0.9355 &  0.0331  &  1.3766  && 1.0155  &  0.0383  &  1.5226 \\
$\beta_{3}$  & 0.6 && 0.7284 & 0.0403 &  1.5750   && 0.5898  & 0.0310 & 1.2686   &&  0.5616  & 0.0372 &  1.2588 \\
$\beta_{4}$  & $-0.8$ && $-0.8699$ & 0.0432 & 1.8581  && $-0.7587$ & 0.0330 & 1.3724  && $-0.8482$   & 0.0369  &  1.2620 \\
$\zeta_{3}$  & $-1.0$ &&  $-1.2082$ & 0.1493 & 1.0653  && $-1.2043$ & 0.1304 & 1.0389  && $-1.0786$ & 0.1451 &  1.0669 \\
$\zeta_{4}$  & 1.0  && 1.2781 &  0.1548 & 1.1919  && 1.0350 & 0.1373 &  1.0649  && 1.1079   & 0.1427  &  1.0793 \\
$\sigma_{\alpha }^{2}$  & 1.0 && 1.1668 & 0.1502 & 2.1466   && 1.1444  &  0.1177 & 1.6006  && 1.1923   & 0.1321 &  1.6553 \\
\hline
& \multicolumn{13}{c}{n=2000} \\
\hline
$\beta_{1}$  & 0.5 &&  0.5241 & 0.0375 & 1.2201  && 0.4812 & 0.0326 & 1.0176 && 0.4661 & 0.0355 &  1.0200 \\
$\beta_{2}$  & 1.0 &&  0.9852 & 0.0281 & 1.6192  && 0.9985 & 0.0249 &  1.6350  &&  0.9784 & 0.0274 & 1.5347 \\
$\beta_{3}$  & 0.6 && 0.6134 & 0.0262 & 1.2643  && 0.5914 & 0.0226 & 1.3154   &&   0.6017 & 0.0259  & 1.3277 \\
$\beta_{4}$  & $-0.8$ && $-0.7745$ & 0.0278 & 1.4142  &&  $-0.7719$ & 0.0235 & 1.4121  && $-0.7897$ & 0.0253 & 1.3079 \\
$\zeta_{3}$  & $-1.0$ &&  $-0.9418$ & 0.0970 & 1.0328  && $-1.0325$ & 0.0894  & 1.0762  && $-1.0957$ & 0.1005 & 1.0553 \\
$\zeta_{4}$  & 1.0 && 0.9678 & 0.0985  & 1.0782  && 1.0814 & 0.0913 & 1.0117 && 1.1127 & 0.0994  & 1.0314 \\
$\sigma_{\alpha }^{2}$  & 1.0 && 0.8584 & 0.0857 & 2.1350  &&  0.9433 & 0.0754 & 1.6048  && 1.0303 & 0.0895 &  1.8290 \\
\hline
\end{tabular}
		\parbox{\linewidth}{\caption{\label{table2}\footnotesize{True values (True), posterior mean (Mean), standard deviation (Std) and inefficiency factor (IF) of the parameters in the simulation study. The upper panel presents results for $n=1,000$ individuals and the lower panel presents results for $n=2,000$ individuals.}}}

%	\parbox{0.95\linewidth}{\caption{\footnotesize{True values (\textsc{true}), posterior mean (\textsc{mean}), standard deviation (\textsc{std}) and inefficiency factor (\textsc{if}) of the parameters in the simulation study. The upper panel presents results for $n=1,000$ individuals and the lower panel presents results for $n=2,000$ individuals.}}}
}
\end{table}
%----------------------------------------------------------------------------------
%----------------- end tables -------------------------

%----------------- end figures  -------------------------
\section{Marginal Effects, Relative Risk and Odds Ratio}\label{sec:ME-RR-OR}

Our proposed binary panel quantile model is nonlinear, as such the
coefficients by themselves do not give the marginal effects
\citep{Rahman-2016,Rahman-Vossmeyer-2019}. However, marginal effects are
important to understand the effect of a covariate on the probability of
success. For example, in our current application one may be interested in
seeing how the probability of recidivism is affected due to an additional
year of schooling, decreasing regional unemployment rate by 1 percentage, or
involvement in violent crime. These may be useful to policy makers and
researchers alike.

To formally derive the marginal effects, we rewrite the BPQRCRE model
presented in
Equation~(1) as follows, %\eqref{eq:model2}
%-----------------------------------
\begin{equation}\label{eq15}
\begin{split}
z_{it} & =  x^{\prime}_{it}\beta + \alpha _{i} + \varepsilon_{it} ,
\qquad \forall \; i=1,\cdots,n,  \; \textrm{and} \; t=1,\cdots,T_i, \\
%-----
\alpha _{i} & \sim N(\overline{m}_{i}^{\prime }\zeta,\sigma_{\alpha }^{2}),
\end{split}
\end{equation}
%----------------------------------
where $\varepsilon_{it} = w_{it}\theta + \tau \sqrt{w_{it}} u_{it}$. We know
$\varepsilon_{it} \overset{iid}{\sim} AL(0,1,p)$ for $i=1,\cdots,n$ and
$t=1,\cdots,T_{i}$, which implies $z_{it}|\alpha_{i} \overset{ind} \sim
AL(x'_{it} \beta + \alpha_{i},1,p)$, where $ind$ denotes independently
distributed.

Given the model framework, the probability of success can be calculated as,
%----------------------------------
\begin{equation}\label{eq:ProbSuccess}
\begin{split}
\Pr(y_{it}  =1| x_{it}, \beta, \alpha_{i}) & =  \Pr(z_{it} > 0| \beta, \alpha_{i}, x_{it}) \\
%-----
&= 1 - \Pr(z_{it} \le 0| \beta, \alpha_{i}, x_{it}) \\
%----
&= 1 - \Pr(\varepsilon_{it} \le -x'_{it} \beta - \alpha_{i}|\beta, \alpha_{i}, x_{it})\\
%----
&= 1 - F_{AL}( -x'_{it} \beta - \alpha_{i},0,1,p),
\end{split}
\end{equation}
%----------------------------------
for $i=1,\cdots,n$ and $t=1,\cdots,T_{i}$, where $F_{AL}(x,0,1,p)$ denotes
the cumulative distribution function (\emph{cdf}) of an AL distribution
evaluated at $x$, with location 0, scale 1 and quantile $p$.

Marginal effect (i.e., the derivative of the probability of success with
respect to a covariate) is often computed at the average covariate values or
by averaging the marginal effects over the sample, \emph{alias} average
partial effects \citep{Wooldridge-2010,Greene-2017}. However,
\citet{Jeliazkov-Vossmeyer-2018} show that both these quantities can be
clearly inadequate in nonlinear settings (e.g., binary, ordinal and Poisson
models) because they employ point estimates rather than their full
distribution. To account for the uncertainty in parameters, we need another
layer of integration over the model parameters. This idea of calculating the
marginal effect that accounts for uncertainty in parameters and the
covariates has been previously considered, amongst others, by
\citet{Chib-Jeliazkov-2006} in the context of semiparametric dynamic binary
longitudinal models, and \citet{Jeliazkov-etal-2008} and
\citet{Jeliazkov-Rahman-2012} in relation to ordinal and binary models.
Within the quantile literature, this has been mentioned by
\citet{Rahman-2016} in the context of ordinal models and discussed by
\citet{Rahman-Vossmeyer-2019} in connection to binary longitudinal outcome
models.

Suppose, we are interested in the average marginal effect i.e., average
difference between probabilities of success when the $j$-th covariate
$\{x_{it,j}\}_{t=1}^{T_{i}}$ is set to the values $a$ and $b$, denoted as
$\{x_{it,j}^{a}\}_{t=1}^{T_{i}}$ and $ \{ x_{it,j}^{b}\}_{t=1}^{T_{i}}$,
respectively. To proceed, we split the covariate and parameter vectors as
follows: $x_{it}^{a} = ( x_{it,j}^{a}, x_{it,-j})$, $x_{it}^{b} = (
x_{it,j}^{b}, x_{it,-j})$, and $\beta = (\beta_{j}, \beta_{-j})$, where $-j$
in the subscript denotes all covariates/parameters except the $j$-th
covariate/parameter. We are interested in the distribution of the difference
$\{\Pr(y_{it}=1|x_{it,j}^{b}) - \Pr(y_{it}=1|x_{it,j}^{a} ) \}$, marginalized
over $\{x_{it,-j}\}$ and the parameters $(\beta,\alpha)$, given the data
$y=(y_{1}, \cdots, y_{n})'$. As done in \citet{Chib-Jeliazkov-2006} and
\citet{Rahman-Vossmeyer-2019}, we marginalize the covariates using their
empirical distribution and integrate the parameters using their posterior
distribution.

To obtain a sample of draws from the distribution of the difference in
probabilities of success, marginalized over $\{x_{it,-j}\}$ and
$(\beta,\alpha)$, we express it as follows,
%------------------------------
\begin{equation}\label{eq:MEdraws}
\begin{split}
& \{\Pr(y_{it}=1|x_{it,j}^{b}) - \Pr(y_{it}=1|x_{it,j}^{a} ) \} \\
& = \int \Big\{ P(y_{it}=1|x_{it,j}^{b}, x_{it,-j}, \beta, \alpha) -
    P(y_{it}=1|x_{it,j}^{a}, x_{it,-j}, \beta, \alpha)  \Big\} \\
    %-----
&   \qquad \times \pi(x_{it,-j}) \pi(\beta|y) \pi(\alpha|y)
    \; d(x_{it,-j})\, d \beta \, d\alpha.
\end{split}
\end{equation}
%------------------------------
Drawing a sample from the above predictive distribution (i.e.,
equation~\ref{eq:MEdraws}) utilizes the method of composition. This
involves randomly drawing an individual, extracting the corresponding
sequence of covariate values, drawing a value $(\beta, \alpha)$ from the
posterior distribution and finally evaluating $\{\Pr(y_{it}=1|x_{it,j}^{b}) -
\Pr(y_{it}=1|x_{it,j}^{a} ) \}$. This is repeated for all other individuals
and other draws from the posterior distribution. Finally, the average
marginal effect $(AME_{Bayes})$ is calculated as the average of the
difference in pointwise probabilities of success as follows,
%-------------------------------------
\begin{equation}\label{eq:AME_Bayes}
\begin{split}
AME_{\text{Bayes}} & \approx \frac{1}{T} \frac{1}{M} \sum_{i=1}^{n} \sum_{t=1}^{T_{i}} \sum_{m=1}^{M}
  \Big[ F_{AL}(-x_{it,j}^{a} \beta_{j}^{(m)} - x'_{it,-j} \beta_{-j}^{(m)}
  - \alpha_{i}^{m},0,1,p)  \\
  %----
& \qquad   - F_{AL}(-x_{it,j}^{b} \beta_{j}^{(m)} - x'_{it,-j} \beta_{-j}^{(m)}
  - \alpha_{i}^{m},0,1,p)    \Big]
\end{split}
\end{equation}
%----------------------------------
where the expression for probability of success follows from
equation~\eqref{eq:ProbSuccess}, $T=\sum_{i=1}^{n} T_{i}$ is the total number
of observations, and  M is the number of MCMC draws. Here, $(\beta^{(m)},
\alpha^{(m)})$ is an MCMC draw of $(\beta, \alpha)$ for $m=1,...,M$. The
quantity in equation~\eqref{eq:AME_Bayes} provides estimate that integrates
out the variability in the sample and the uncertainty in parameter
estimation.

Relative risk ($RR$) can be calculated to demonstrate the association between
the risk factor or exposure ($x_j$) and the event ($y$) being studied. It is
the ratio of the probability of the outcome with the risk factor ($x_j=b$) to
the probability of the outcome with the risk factor ($x_j=a$) (\textit{e.g.},
exposed ($b=1$) /non-exposed ($a=0$)). Following
equation~\eqref{eq:AME_Bayes}, the relative risk is given by,
%---------------------------------
\begin{equation}
RR(b/a)_{\text{Bayes}}=  \frac{1}{T}  \frac{1}{M} \sum\limits_{i=1}^{n}
\sum\limits_{t=1}^{T_{i}}  \sum\limits_{m=1}^{M}
\frac{H^{b}_{AL} }{H^{a}_{AL} }.
\label{eq:RR}
\end{equation}%
%---------------------------------
where $H_{AL}^{r} = 1 - F_{AL}(-x_{it,j}^{r} \beta_{j}^{(m)} - x'_{it,-j}
\beta_{-j}^{(m)} - \alpha_{i}^{m},0,1,p)$ for $r = a, b$, is the complement
of the \emph{cdf} of the AL distribution. If there is a causal effect between the
exposure and the outcome, values of $RR$ can be interpreted as follows: if
$RR>1$ (resp. $RR<1$), the risk of outcome is increased (resp. decreased) by
the exposure and if $RR=1$, the exposure does not affect the outcome.

The odds ratio is the ratio of the odds of the event occurring with the risk
factor ($x_j=b$) to the odds of it occurring with the risk factor ($x_j=a$).
It is given by:
%----------------------------------
\begin{equation}
OR(b/a)_{\text{Bayes}}= \frac{1}{T}  \frac{1}{M} \sum\limits_{i=1}^{n}
\sum\limits_{t=1}^{T_{i}}  \sum\limits_{m=1}^{M}
\left( \frac{ H^{b}_{AL}  }{ 1 - H^{b}_{AL}   }  \right) \bigg/
\left( \frac{H^{a}_{AL}   }{   1 - H^{a}_{AL}  }  \right).
\label{eq22}
\end{equation}%
%----------------------------------
The odds ratio, for a given exposure $x_j$, does not have an intuitive
interpretation as the relative risk. OR are often interpreted as if they were
equivalent to relative risks while ignoring their meaning as a ratio of odds.
Two main factors influence the discrepancies between $RR$ and $OR$: the
initial risk of an event $y_{it}$, and the strength of the association
between exposure $x_{it,j}$ and the event $y_{it}$. When the event $y_{it}=1$
is rare, then $OR(b/a)$ $\approx $ $RR(b/a)$, but the odds ratio generally
overestimates the relative risk, and this overestimation becomes larger with
increasing incidence of the outcome.

%-------------------------------------------------------------------------------------------
%-------------------------------------------------------------------------------------------

\section{An application to crime recidivism in Canada}
\label{sec:application} Crime has been extensively studied by economists both
theoretically and empirically (see, \textit{e.g.}\cite{chalfin2017} for a
recent survey). Many empirical analyses have used panel data either at the
state  \citep{Cornwell1994,baltagi2006,baltagi2017} or at the individual
level \citep{bhuller2019}. The vast majority of the published papers focus on
the situation in the U.S. Here, we study crime recidivism in Canada between
2007-2017 for two reasons.  First, the Canadian government implemented a
``tough-on-crime" policy in 2012 which  marked a shift from rehabilitating to
warehousing people. Our proposed estimator is well suited to measure the
sensitivity of recidivism to this new policy.\footnote{Starting in 2012, the
government enacted a series of legislations that made  prison conditions more
austere; imposed lengthier incarceration periods; significantly expanded the
scope of mandatory minimum penalties; and reduced opportunities for
conditional release, parole, and alternatives to incarceration.} Second,
offenders who are sentenced to less than two years serve their sentence in a
provincial \textit{correctional institution} while offenders sentenced to two
years or more  serve their's  in a \textit{federal penitentiary}. The former
have committed less serious crimes and are more likely to reoffend over the
time span of our panel. Because our analysis focuses on this population,  the
impact of the ``tough-on-crime'' policy may be more easily unearthed from the
data than if it  focused on detainees serving long sentences.

\subsection{The data}
\label{subsec:data}

We utilize a sample data drawn from the administrative correctional files for
the Province of Quebec. The files are used by corrections personnel to manage
activities and interventions related to housing offenders and contain
detailed information on inmates' characteristics, correctional facilities,
and sentence administration. While they offer a wealth of information, the
files have never been used for research purposes.   For illustrative purpose,
we have drawn a random sample of 8,974 detainees out of a population of
148,441. Each detainee is observed upon release and up until 2017. The
earliest releases occur in 2007 and the latest in 2016. Overall, our
unbalanced panel includes 61,880 observations. Of the 8,974 detainees, as
many as 3,466 had at least one repeat offense over our sample period.

%--------------------------------------------------------------------------------------
\begin{table}
		\captionsetup{justification=centering}	\centering
	\begin{tabular}{lr@{.}lr@{.}l}
		\toprule
		& \multicolumn{2}{c}{Mean} & \multicolumn{2}{c}{Std}  \\
		\midrule
		Age & 41&366 & 12&596 \\
		Schooling & 6&011 & 3&814\\
		Married & 0&045 & 0&208  \\
		 Aboriginal$^\dagger$  & 0&045& 0&206  \\
		Mother Tongue Not Fr. or Eng. & 0&070 & 0&255  \\
	    Type of Crime:\\
	    ~~~Traffic Related & 0 &163 & 0&384\\
		~~~Violent (Domestic, Assault \& Battery, \textit{etc.})& 0&099 & 0&299 \\
		~~~Property (Theft, Robbery, \textit{etc.}) & 0&439 & 0&496 \\
		~~~Other Infractions to Criminal Code& 0&299 & 0&458 \\
		Unemployment rate & 8&329 & 2&063  \\
%		Year 2012 & 0&742 & 0&438\\

%		pre_period & 0.0000 & 0.0000 & 0.0000 & 0.0000 \\
%		\hline
%		post_period & 0.2516 & 0.4339 & 0.0000 & 1.0000 \\
%		\hline
%		pre_post_period & 0.7484 & 0.4339 & 0.0000 & 1.0000 \\
%		\hline
%		pre_post_period_0 & 0.2585 & 0.4378 & 0.0000 & 1.0000 \\
%		\hline
%		pre_post_period_1 & 0.4899 & 0.4999 & 0.0000 & 1.0000 \\
%%		\hline
%		recidivism_pre_period & 0.0000 & 0.0000 & 0.0000 & 0.0000 \\
%		\hline
%%		recidivism_post_period & 0.0229 & 0.1497 & 0.0000 & 1.0000 \\
%		\hline
%		recidivism_pre_post_period & 0.0909 & 0.2875 & 0.0000 & 1.0000 \\
%		\hline
        Post 2012 (=1) & 0&252 & 0&434\\
		Recidivism Entire Sample & 0&114 & 0&318 \\
		Recidivism Pre-Post 2012 & 0&091 & 0&288  \\
		Recidivism Post 2012 & 0&023 & 0&150 \\
%		\hline
%		Demeaned Age & -0.0000 & 1.2596 & -2.4366 & 5.3634 \\
%		\hline
%		Demeaned Schooling & 0.0000 & 3.8144 & -6.0113 & 16.9887 \\
%		\hline
%		Demeaned Unemployment rate & 0.0000 & 2.0628 & -3.9286 & 9.1714 \\
%		\hline
%		xbar_Age & -0.0000 & 1.2745 & -2.1414 & 4.9086 \\
%		\hline
%		xbar_Schooling & -0.0000 & 3.6505 & -5.6910 & 17.3090 \\
%		\hline
%		xbar_Unemp & -0.0000 & 1.9434 & -3.6486 & 7.2105 \\
		\bottomrule
$^\dagger$ First Nations, Inuit and M\'etis.
	\end{tabular}
		\caption{Descriptive Summary of the Sample Data.}
\label{table:Descriptive}
\end{table}
%--------------------------------------------------------------------------------------

Table \ref{table:Descriptive} presents the main characteristics of our
sample. Detainees are 41 years of age on average, have a level of schooling
corresponding to a high-school degree, and few are married.   Aboriginal
detainees represent 4.5\% of our sample  and most are incarcerated in a
correctional institution suited to their needs and specificities.
Approximately 7\% of inmates do not have French or English, Canada's two
official languages, as their  mother tongue.  These include some Aboriginal
residents as well as recent immigrants.  Crimes have been aggregated into 4
distinct categories.  By far the most common concerns property crime. Traffic
related and infractions to the criminal code usually entail shorter
sentences. Violent crimes receive the longest sentences in our data but
necessarily less than two years. As mentioned above, major crimes fall under
the federal jurisdiction. The yearly unemployment rate is measured at the
regional level where a detainee is released. Over our sample period, it
varies between 4.4\% and 17.5\%.  The ``Post 2012'' variable is equal to one
if a detainee entered the panel at any time during or after 2012 while the
``Pre-Post 2012'' variable is equal to one if a detainee entered at any time
before  2012. In the latter case, repeat offenses  are observed over the
entire duration of the panel, i.e. 2007-2017.  In the former, they are only
observed over 2012-2017. Roughly a quarter of our sample belongs to the
period post the implementation of ``tough-on-crime'' policy.  The remaining
observations  (74.8\%) were sanctioned  prior to 2012 and may or may not have
reoffended in the Post 2012 period. The next 3 lines of the table provide
information on the rates of recidivism for distinct
periods.\footnote{Recidivism is a yearly dummy variable equal to one the year
at which the new incarceration begins and zero otherwise. Recidivism may be
equal to one in consecutive years so long as the repeat offenses occurred
after the end of the previous sentence.  Reincarcerations while on parole or
on conditional release are not considered repeat offenses.  }  Thus, the
overall rate of recidivism is equal to 11.4\%. The next line focuses on
individuals who are present both before and after the implementation of the
``tough-on-crime'' policy. Their recidivism rate is approximately 9\%.  The
last line focuses on individuals who entered the panel on or after 2012.
Naturally, as they are observed for a shorter period of time, their
recidivism rate is relatively smaller at 2.3\%.

\begin{figure}
	\centering
	\captionsetup{justification=centering}
	\includegraphics[width=0.7\linewidth]{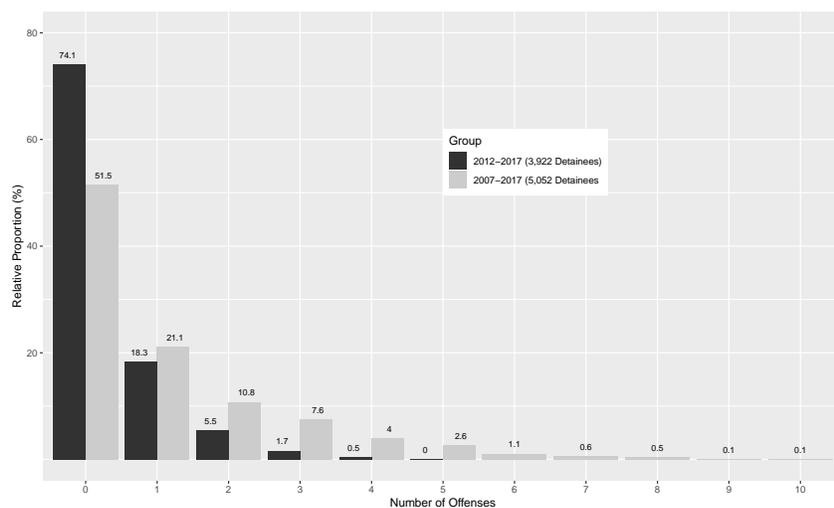}
	\parbox{\linewidth}{\caption{\footnotesize{Frequency of Repeat Offenses}	\label{fig:Repeat}}}
\end{figure}

Figure \ref{fig:Repeat} depicts the proportions of repeat offenses for the
entire sample period and for those who entered the panel in 2012 or later.
The figure provides \textit{prima facie} evidence on the impact of the
policy.  Indeed, the proportion of detainees who do not reoffend upon release
in the post-policy period is 15 percentage points larger (74.1\%) than the
proportion for the whole sample period (51.5\%).  Likewise, the proportion of
repeat offenders is between 3 to 6 percentage points lower in the post-policy
period for any given number of repeat offenses.\footnote{Obviously, detainees
who entered the sample on or after 2012 have had less time to reoffend.  Yet,
in our sample as many as 34\% of detainees are reincarcerated within 12
months upon release, and as many as 43\% within two years. Hence, the sharp
decline in repeat offenses in the post-2012 period is unlikely due to the
sampling frame. See \cite{lalande2015}.} Naturally, such differences may
results from factors other than the ``tough-on-crime"  policy, such as, but
not limited to, better economic opportunities, and demographic compositional
changes. In order to net  these out, we now turn to formal econometric
modelling.\footnote{To the extent the new legislation has indeed lowered the
recidivism rates, it not clear whether it did so through deterrent or
incapacitative effects. Yet, see \cite{bhuller2019} for U.S. evidence
according to which deterrence dominates incapacitation.}

%----------------------------------------------------------------------------------------
\vskip-220pt
%----------------------------------------------------------------------------------------

\subsection{Estimation results}
The dependent variable $y$ is an indicator variable that equals 1 if an
individual commits a repeat offense and 0 otherwise. We regress the
probability of recidivism on time-varying covariates (age, schooling,
unemployment rate), on time-invariant policy variables (\texttt{Pre-Post
2012} and \texttt{Post 2012}) and on other time-invariant control
variables.\\  Our Bayesian setup uses the same independent prior
distributions as in the simulation exercise: $\beta \sim N_{k}\left( 0_{k}  ,
10^3 I_{k} \right)$, $\zeta \sim N_{k-1}\left( 0_{k-1} , 10^3 I_{k-2}
\right)$, $\sigma_{\alpha }^{2} \sim IG \left( 10/2 , 9/2 \right)$. We
generate $60,000$ MCMC samples of which the first $10,000$ are discarded as
burn-ins. The posterior estimates are reported using a thinning factor of
$50$, optimized following the approach in
\citet{Owen-2017}.\footnote{Thinning has been criticized by some
\citep{MacEachern-Berliner-1994,Link-Eaton-2012} while others acknowledge
that it  can increase statistical efficiency \citep{Geyer-1991}. See
\cite{Owen-2017} who claims that the arguments against thinning may be
misleading.}

%----------------------------------------------------------------------------------------------
\begin{figure}[b!]
\centerline{
\makebox{\includegraphics[width=6.75in, height = 8.5in, trim = {0 1.5cm 0 1.4cm}, clip]{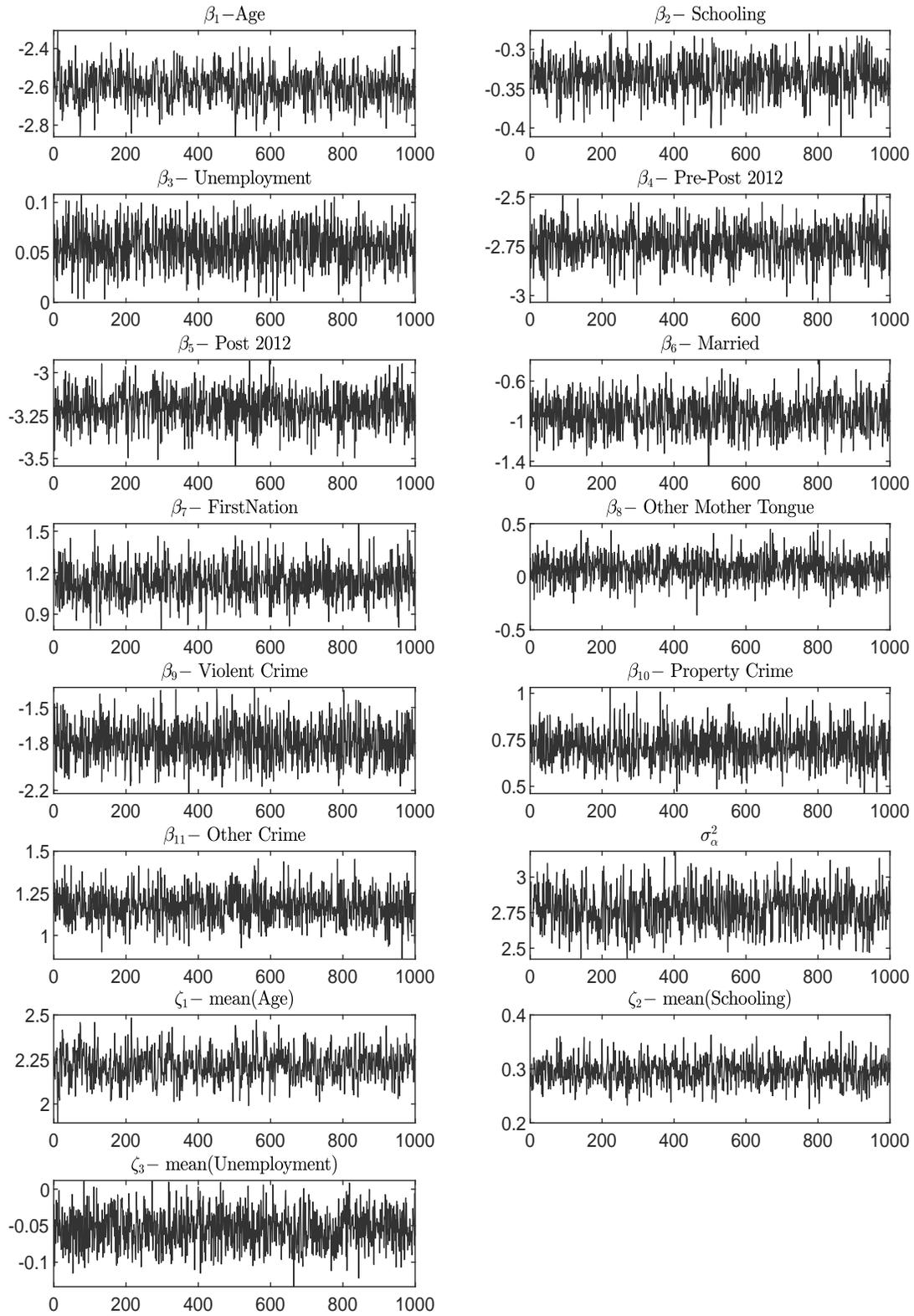} } }
\caption{Trace plots of the parameters for the 75th quantile.}
\label{fig:CrimeTrace-75th}
\end{figure}
%----------------------------------------------------------------------------------------------

%----------------------------------------------------------------------------------------------
\begin{figure}[t!]
\centerline{
\makebox{\includegraphics[width=6.5in, height = 4.0in, trim = {0 0 0 0}, clip]{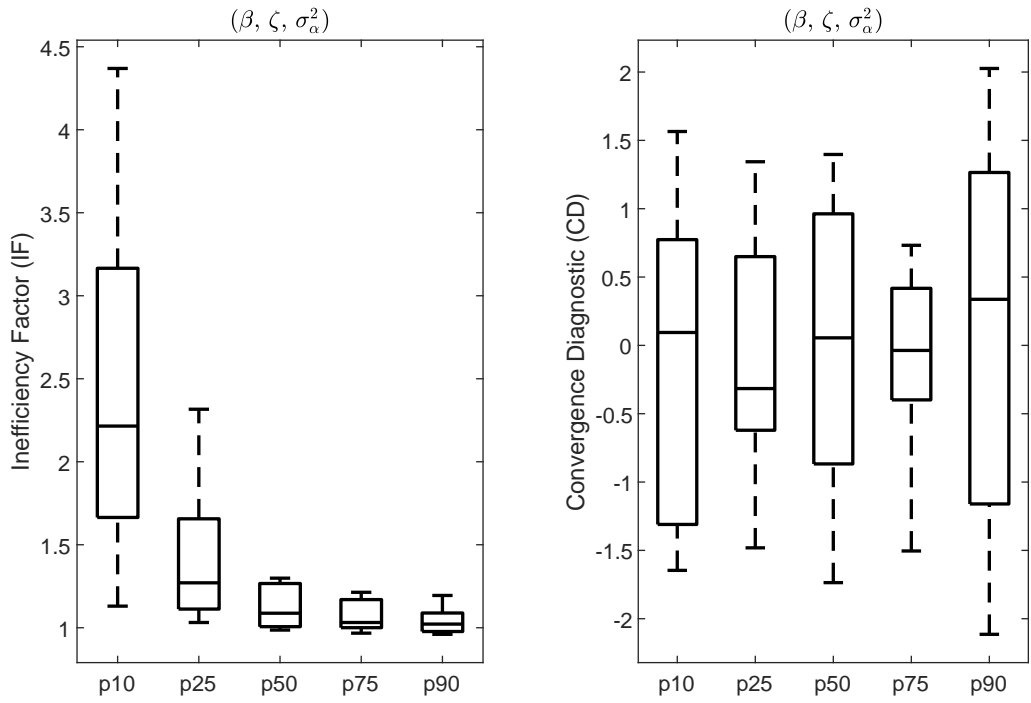} } }
\caption{Boxplots of the Inefficiency Factors and Convergence Diagnostics for ($\beta, \zeta, \sigma_{\alpha}^{2}$) at 5 different quantiles.}
\label{fig:CrimeBoxplots}
\end{figure}
%----------------------------------------------------------------------------------------------

The mixing of the MCMC chain is extremely good as illustrated in
Figure~\ref{fig:CrimeTrace-75th} which exhibits the trace plots  of the
parameters at the 75th quantile.\footnote{Note that the time-varying
covariates (\texttt{Age},\texttt{ Schooling} and\texttt{ Unemployment rate})
have been ``demeaned'' and that \texttt{Age} has been divided by 10. The
parameter estimates must thus be interpreted accordingly.} Trace plots at
other quantiles are similar and not reported for the sake of brevity but they
are available upon request. Figure \ref{fig:CrimeBoxplots} provides
additional information on the performance of the MCMC chain.  The
left-hand-side figure depicts the boxplots of the inefficiency factors of the
parameters ($\beta$s, $\zeta$s and $\sigma^2_{\alpha}$) for each of the five
different quantiles used in estimating the model. Except perhaps for the 10th
quantile, all are reasonably close to one. Consistent with the simulation
results, the parameter with the largest inefficiency factor at the 10th
quantile is $\sigma^2_{\alpha}$ (not shown, see Table \ref{table2}). The
right-hand-side figure reports the boxplots of the  convergence diagnostics
of the parameter estimates for the same five specifications based on the
first 10\% and the last 40\% values of the Markov chain \citep{geweke1992}.
As depicted, all  parameters have $Z$-scores within 2 standard deviation of
the mean at the $5\%$ level or within $2.58$ standard deviation at $1\%$
level.  All in all, the Markov chains behave satisfactorily and thus lend
themselves to statistical inference.

Table \ref{table:Estimates} reports the posterior means and standard
deviations at five different quantiles separately. To ease interpretation,
the quantile-specific estimates are reported column-wise in increasing order.
Row-wise, we distinguish the time-varying covariates from the time-invariant
and the correlated random effects variables. Note that the correlated random
effects specification does not include an intercept. This is to allow the
identification of the two time-invariant policy variables, \texttt{Pre-Post
2012} and \texttt{Post 2012}.  The former, is equal to one if the detainee
was incarcerated prior to 2012 and thus observed both before and after the
implementation of the ``tough-on-crime'' policy.  The latter is equal to one
if a detainee's  first incarceration occurred during or after 2012, and thus
always exposed to the policy.   All other time-invariant variables are
measured at first entry in the panel.\footnote{Recall from Table
\ref{table:Descriptive} that very few men are married.  In addition, next to
none report a change in their marital status in between incarcerations.
Further,  since the marital status of non-repeaters is not observed in the
data we are constrained to use the information at entry in the panel.} The
estimates of the correlated random components associated with the individual
mean \texttt{Age}, \texttt{Schooling} and \texttt{Unemployment},
$\widehat{\zeta}$,  are all statistically different from zero regardless of
the quantile. The individual-specific effects, $\alpha_i$, are thus highly
correlated with the individual means of the time-varying variables. Omitting
this correlation may therefore bias the model estimates and hence their
intrinsic marginal effects and relative risks. This provides empirical
support to the worthiness of incorporating correlated random effects within a
quantile regression.

%-----------------------------------------------------------------------------------------
\begin{table}[t!]
\centering \footnotesize \setlength{\tabcolsep}{3pt}
\setlength{\extrarowheight}{2pt}
\begin{tabular}{l rrr rrr rrr rrr rrr r}  %{lr@{.}lr@{.}lr@{.}lr@{.}lr@{.}lr@{.}lr@{.}lr@{.}lr@{.}lr@{.}lr@{.}l}
\toprule
Variable && \multicolumn{2}{c}{$p=10\%$} && \multicolumn{2}{c}{$p=25\%$}
         && \multicolumn{2}{c}{$p=50\%$} && \multicolumn{2}{c}{$p=75\%$}
         && \multicolumn{2}{c}{$p=90\%$} & \\
\cmidrule{3-4} \cmidrule{6-7} \cmidrule{9-10} \cmidrule{12-13} \cmidrule{15-16}
&& \multicolumn{1}{c}{Mean} & \multicolumn{1}{c}{Std} && \multicolumn{1}{c}{Mean} & \multicolumn{1}{c}{Std}
&& \multicolumn{1}{c}{Mean} & \multicolumn{1}{c}{Std} && \multicolumn{1}{c}{Mean} & \multicolumn{1}{c}{Std}
&& \multicolumn{1}{c}{Mean} & \multicolumn{1}{c}{Std} & \\
\midrule
%------------------------------------------------------------------------------
\multicolumn{16}{l}{Time varying covariates}\\
\hspace{0.15in} $\beta$-Age	&&	$-12.634$	& 0.426 && $-5.417$ &	0.183	&&	$-3.024$ & 0.099
                            &&	$ -2.598$   & 0.083	&& $-3.829$ &   0.123 &	\\
\hspace{0.15in} $\beta$-Schooling	
                && $-1.474$ & 0.116 && $-0.649$	& 0.047	&& $-0.379$	&  0.025
                && $-0.335$	& 0.021	&& $-0.484$	& 0.032	&	\\
\hspace{0.15in} $\beta$-Unemp Rate	
                && 0.347 &	0.103 && 0.140	& 0.040	&& 0.072 &	0.022	
                && 0.056 &	0.020 && 0.078	&	0.028	&	\\
%------------------------------------------------------------------------------
\multicolumn{16}{l}{Policy Variables (Time invariant)}\\
\hspace{0.15in} Pre-Post 2012	
                &&	$-28.671$ &	0.465  && $-11.299$	& 0.203	 &&	$-5.034$ & 0.113		
                &&	$-2.735$  &	0.087  && $-1.800$	&	0.121	&	\\
\hspace{0.15in} Post-2012	
                &&	$-31.238$ &	0.522  && $-12.389$	&  0.224 &&	$-5.610$ &	0.125		
                &&	$-3.202$  &	0.097  && $-2.504$	&  0.134	&	\\
\multicolumn{16}{l}{Other Time invariant covariates}\\
%------------------------------------------------------------------------------
\hspace{0.15in} Married	
                &&	$-5.315$  &	0.901  && $-2.286$	& 0.399	 &&	$-1.222$ &	0.204	
                &&	$-0.932$  &	0.164  && $-1.298$	& 0.226	 &	\\
\hspace{0.15in} Aboriginals	
                &&	$5.661$	  &	0.634  && $2.494$	& 0.295	 &&	$1.359$	 &	0.148		
                &&	$1.132$	  &	0.121  && $1.645$	& 0.197	 &	\\
\hspace{0.15in} Oth. Mot. Ton.	
                &&	$0.329$	  &	0.683  && $0.174$	& 0.270	 &&	0.090 &	 0.147		
                &&	0.080	  &	0.120  && $0.089$	& 0.164	 &	\\
%------------------------------------------------------------------------------
\hspace{0.15in} Violent Crime	
                &&	$-12.286$ &	0.948  && $-4.908$	& 0.419	 &&	$-2.484$ & 0.224		
                &&	$-1.797$  &	0.159  && $-2.253$	& 0.205	 &	\\
\hspace{0.15in} Property Crime	
                &&	$3.111$	  &	0.495  && $1.674$	& 0.210	 &&	$0.906$	 & 0.112	
                &&	$0.713$	  &	0.090  && $0.967$	& 0.126	 &	\\
\hspace{0.15in} Other Crime	
                &&	$5.656$	  &	0.498  && $2.707$	& 0.212	 &&	$1.456$	 & 0.116	
                &&	$1.171$	  &	0.092  && $1.636$	& 0.131	 &	\\
%------------------------------------------------------------------------------
\multicolumn{16}{l}{Correlated Random Effects}\\
\hspace{0.15in} $\zeta$-Age	
                &&	$10.381$  &	0.456	&&	$4.521$	& 0.195	 &&	 $2.558$   & 0.105		
                &&	$ 2.218$  &	0.085	&&	3.286	& 0.123	 &\\
\hspace{0.15in} $\zeta$-Schooling	
                &&	$1.263$	  &	0.127	&&	$0.560$	& 0.051	 &&	 $0.331$   & 0.027		
                &&	$0.295$	  &	0.023	&&	$0.426$	& 0.034	 &	\\							
\hspace{0.15in} $\zeta$-Unemployment	
                &&	$-0.311$  &	0.137	&&	$-0.129$ & 0.054 &&	 $-0.067$  & 0.029		
                &&	$-0.054$  &	0.025	&&	$-0.080$ & 0.037 &	\\			
\hspace{0.15in} $\sigma^2_\alpha$	
                &&	$75.810$  &	3.325	&&	$13.133$ & 0.540 &&	 $3.871$   & 0.165		
                &&	$2.777$	  &	0.132	&&	$6.217$	 & 0.325 &	\\
\bottomrule
\end{tabular}
%------------------------------------------------------------------------------
\parbox{\linewidth}{\caption{\small{Posterior Mean (Mean) and Standard Deviation (Std)
                    of the Parameters in the Crime Application.}}
\label{table:Estimates}
}
\end{table}
%-----------------------------------------------------------------------------------------

The first noteworthy feature of the table is that all parameter estimates are
statistically different from zero, except for the parameter associated with
\texttt{Other Mother Tongue}.  Thus detainees who report speaking a language
other than English or French at home are no more and no less likely to
eventually reoffend.  A second interesting feature concerns the sign of the
parameter estimates. Indeed, all are consistent with recent research on crime
recidivism.  For instance,  \texttt{Age} and \texttt{Schooling} are
associated with lower rates of recidivism \citep{bhuller2019} whereas being
released during a period of high unemployment has been found to favour
recidivism \citep{siwach2018,rege2019}. Likewise, married men are less likely
to reoffend whereas  Aboriginal detainees are more likely to do so
\citep{justice2017}. The type of crime is also associated with recidivism.
The estimates must be interpreted relative to traffic related crimes, which
is the base or omitted category in our analysis. Clearly, sentences for
\texttt{Violent Crimes} will be harsher and so the large parameter estimate
presumably reflects an incapacitative effect. Finally, the parameter
estimates of  \texttt{Post 2012} is larger than that of \texttt{Pre-Post
2012} which suggests that the implementation of the ``tough-on-crime'' policy
may have had a detrimental effect on recidivism.

%-------------------------------------------------------------------------
\begin{figure}[h!]
\centerline{
\makebox{\includegraphics[width=6.75in, height = 6.5in, trim = {0 1.2cm 0 1cm}, clip]{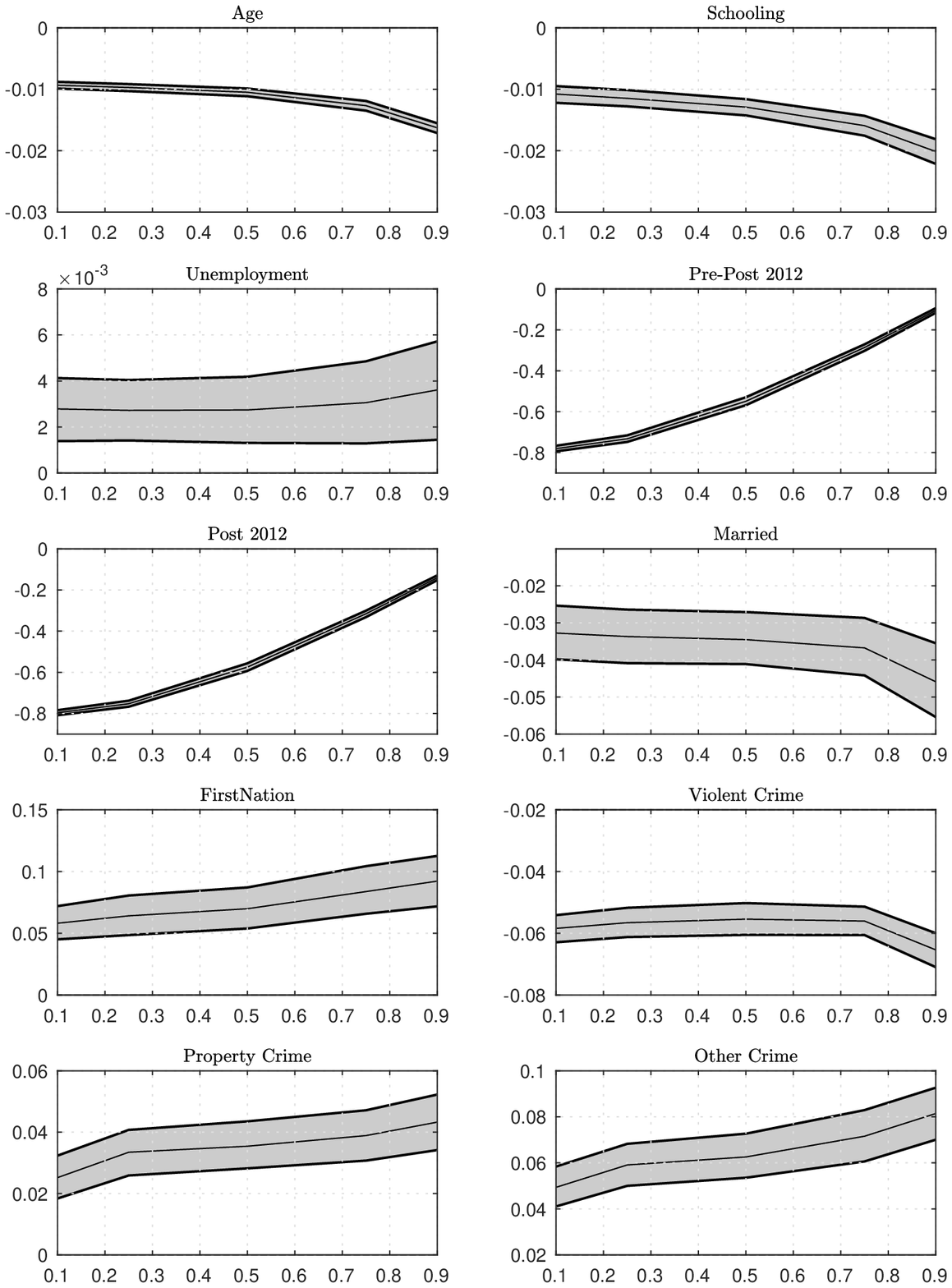} } }  % trim ={left,bottom,right,top}
\caption{Marginal Effects with 95\% HPDI.}
\label{fig:MarginalEffects}
\end{figure}
%-------------------------------------------------------------------------

As stated in Section~\ref{sec:ME-RR-OR}, the parameter estimates such as
those reported in Table \ref{table:Estimates} do not give the marginal
effects. Yet, the latter are important from a policy perspective. Thus, while
the parameter estimates vary considerably across quantiles, it is not clear
that the marginal effects are equally sensitive since they depend both on the
time-varying variables and the correlated random components. Figure
\ref{fig:MarginalEffects} reports the average  marginal effects computed
according to equation (\ref{eq:AME_Bayes}), along with their highest
posterior density intervals (HPDI).\footnote{The marginal effects for
\texttt{Age} correspond to 1/10 of an additional year relative to the mean.
Those for \texttt{Unemployment} and \texttt{Schooling} correspond to one
additional year and one additional percentage point relative to their
individual means, respectively.  The remaining marginal effects  correspond
to a change in the indicator variables.} Note that most marginal effects have
a relatively flat profile between $p10$ and $p75$ and then exhibit a small
kink between $p75$ and $p90$. For instance,  increasing \texttt{Age} by
1/10th reduces the probability of reoffending by 1\% at the 10th quantile and
by 1.6\% at the 90th quantile. Similar results hold for \texttt{Schooling}
(1\% \textit{vs} 2.0\%), \texttt{Married} (0.3\%  \textit{vs}  0.45\%), and
\texttt{Violent Crime} (5\%  \textit{vs} 6.5\%). Thus, for all three
time-varying covariates  the marginal effects increase by one half as we move
from $p10$ to $p90$.  As for the time-invariant variables, their marginal
effects all increase by at least 50\% as we move from  $p10$ to $p90$.  In
particular, the marginal effects associated to \texttt{First Nation},
\texttt{Property Crime} and \texttt{Other Crime} exhibit a twofold increase.
More importantly,  the marginal effects of the two ``tough-on-crime''
variables increase manifold and in a steady fashion between $p10$ and $p90$.
Furthermore, the HPDI is relatively narrow in both cases. Hence,  according
to the parameter estimates associated with \texttt{Pre-Post 2012}, the
probability of reoffending decreases from 78\% at the 10th quantile to as
little as  10\% at the 90th. Likewise, the parameters of \texttt{Post 2012}
imply that the probability decreases from  79\% to 14\% at both extremes.
These results are important from a policy perspective for two reasons. First,
they imply that  detainees from both groups are sensitive to the
``tough-on-crime'' policy, and even more so for those in the \texttt{Post
2012} group.  Consequently long-run recidivism  (\textit{i.e.} recidivism by
the \texttt{Pre-Post 2012} group between 2012-2017) can be addressed just as
well as short-run recidivism (\textit{i.e.} recidivism by the \texttt{Post
2012} group between 2012-2017) by such policies.  Second, the policy does not
impact all detainees alike. Those in the lower quantiles are much more
responsive than those in the upper quantiles.

%-------------------------------------------------------------------------------------
\begin{figure}[t!]
\centerline{
\makebox{\includegraphics[width=6.75in, height = 6.5in, trim = {0 1.2cm 0 1cm}, clip]{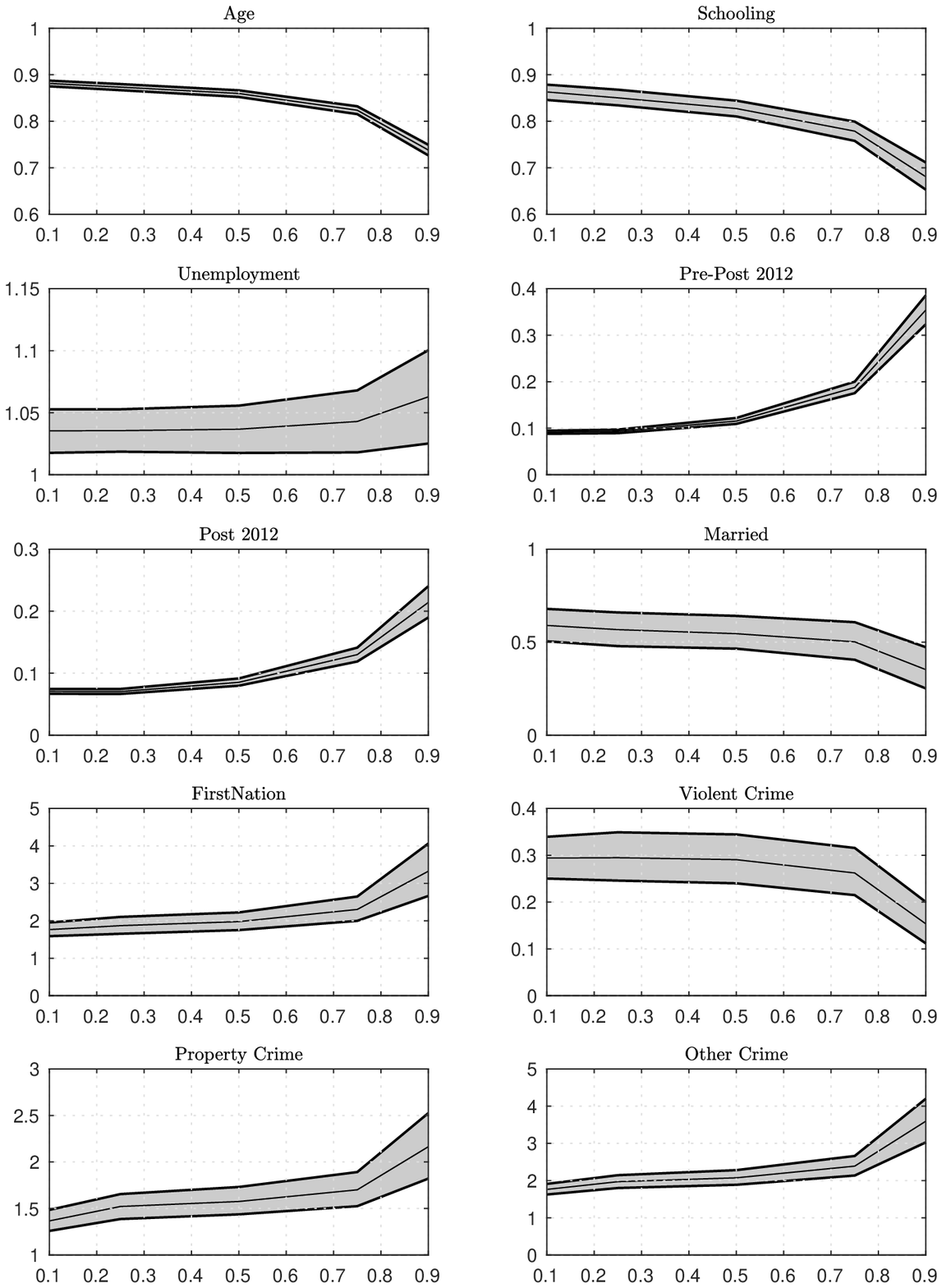} } }  % trim ={left,bottom,right,top}
\caption{Relative Risks with 95\% HPDI.}
\label{fig:CrimeRelativeRisk}
\end{figure}
%-------------------------------------------------------------------------------------

In order to gain further insight into the sensitivity of recidivism to
various covariates, we report the corresponding relative risks in
Figure~\ref{fig:CrimeRelativeRisk}  (see equation (\ref{eq:RR})) along with
their HDPI. Not surprisingly given the marginal effects, the relative risks
are fairly  constant for the first two or three quantiles ($p=10\%,25\%,
50\%$), with a few exceptions. Beyond the second or third quantiles, most
increase or decrease sharply. The figure also shows which covariates
influence recidivism most.  Thus, while \texttt{Age}, \texttt{Schooling} and
\texttt{Unemployment Rate} are associated with slightly different rates of
repeat offenses, only those in the highest quantiles exhibit significantly
different recidivism rates. On the other hand, marital status
(\texttt{Married}), \texttt{First Nation} and types of crime
(\texttt{Violent, Property, Other}) all have significantly higher or lower
relative risks of reoffending as the case may be, and all exhibit a sharp
change between the last two quantiles.  Here, as with the previous figure,
the results concerning the ``tough-on-crime''  variables are particularly
interesting.  Indeed, according to the figure all detainees were much less
likely to reoffend in the post 2012 period, irrespective of whether they
where first convicted prior to  2012 or after. As with the marginal effects,
the policy appears to have had  a larger impact on those in the lower
quantiles. Thus for every quantile the risk of recidivism is much lower (and
significantly different) for those who were exposed to the ``tough-on-crime''
policy. For instance, the 95\% HPDI at quantile $p10$ is
$\left[0.087;0.094\right]$ for the \texttt{Pre-Post 2012} group and
$\left[0.066; 0.074\right]$ for the \texttt{Post 2012} group.  On the other
hand, the 95\% HPDI at quantile $p90$ for the two groups  are
$\left[0.323;0.385\right]$ and  $\left[0.189; 0.240\right]$, respectively. In
other words, for the lowest quantile ($p10$), exposure to the policy
decreases recidivism by as much as $\left[90;91\right]\%$ and
$\left[92;93\right]\%$ for the \texttt{Pre-Post 2012} and \texttt{Post 2012}
groups, respectively. In contrast, for those in the highest quantile, $p90$,
the \texttt{Post 2012} group decreases its recidivism rate more than that of
the \texttt{Pre-Post 2012} ($\left[76;81\right]\%$ vs
$\left[61;67\right]\%$).

\section{Conclusion}\label{sec:conclusion}

This paper presents a panel quantile regression model for binary outcomes
with correlated random-effects (CRE) and proposes two MCMC algorithms for its
estimation. By incorporating the CRE into the panel quantile regression for
discrete outcomes, we move beyond the random-effects framework typically
considered in the Bayesian quantile regression literature. The paper makes an
important contribution to the literature on quantile regression for panel
data and panel quantile regression for discrete outcomes. The two proposed
MCMC algorithms are simpler to implement, but we prefer the algorithm that
exploits block sampling of parameters to reduce the autocorrelation in MCMC
draws. This blocked algorithm is tested in multiple simulation studies and
shown to perform extremely well. We also emphasize the calculation of
marginal effects in models with discrete outcome and explain its computation,
along with those of relative risk and odds ratio, using the MCMC draws.
Finally, we implement the proposed quantile framework to analyze crime
recidivism in Quebec (a Canadian Province) for the period 2007$-$2017 using a
novel data from the administrative correctional files. Amongst other things,
we investigate the effect of the recently implemented ``tough-on-crime''
policy on the probability of repeat offense. Our results show that the policy
negatively affects the probability of repeat offenses across quantiles and
hence has been largely successful in achieving its objective. Besides, the
results suggest that the CRE structure is relevant in modeling the
probability of repeat offenses across quantiles.

This paper opens avenues for future research in several directions. The
proposed framework can be readily extended to panel quantile regression
models with continuous and other discrete response variables (e.g., count and
ordinal outcomes). One may also consider the Hausman-Taylor version of CRE,
where the individual-specific effects are related to only some of the
time-varying and time-invariant regressors, and merge it with the panel
quantile regression model for continuous or discrete outcomes. Besides, a
dynamic relationship can be introduced to panel quantile regression models
(with continuous or discrete outcomes) and the initial condition problems can
be tackled using the CRE structure.

%\begin{acknowledgements}
%This paper is written in honor of Professor Badi H. Baltagi. We are grateful to Bernard Ch\'en\'e, Senior Advisor, Programs Directorate, Public Safety (Qu\'ebec), for his advice and for granting us  access to the data used in the paper. We are also grateful to William Arbour and Steeve Marchand for their advice and numerous discussions. The usual disclaimers apply.
%\end{acknowledgements}

\section*{Conflict of interest}
The authors declare that they have no conflict of interest.


\bibliographystyle{spbasic}
\bibliography{Manuscript1}

\end{document}

\title{Supplementary material - Bayesian quantile regression for a binary outcome model with correlated random effects}
\subtitle{An application on crime recidivism in Canada}

\author{Georges Bresson \and Guy Lacroix \and Mohammad Arshad Rahman}

\institute{Georges Bresson \at
              Department of Economics, Universit\'{e} Paris II, Paris, France.\\
             \email{georges.bresson@u-paris2.fr} \\
             \emph{Corresponding author. Department of Economics, Universit\'{e} Paris II, 12 place du Panth\'{e}on, 75231 Paris cedex 05, France (Tel.: +33 (1) 44 41 89 73).}  %  if needed
           \and
           Guy Lacroix \at
           Department of Economics, Universit\'{e} Laval, Quebec, Canada.\\
           \email{guy.lacroix@hec.ca}
            \and
           Mohammad Arshad Rahman \at
           Department of Economic Sciences, Indian Institute of Technology, Kanpur, India.\\
           \email{marshad@iitk.ac.in}
}

\date{January 2020}

\maketitle

\tableofcontents

\newpage

\section{Conditional densities for the non-blocked sampling in the BPQRCRE model}
\label{sec:nonblocked}

The binary panel quantile regression with correlated random effects (BPQRCRE) model, when stacked for individual $i$, can be written in the hierarchical form as follows,
%-----------------------------------
\begin{equation}
\begin{split}
z_{i} & =  X_{i}\beta +\iota _{T_{i}}\alpha _{i}+ w_{i} \theta + D_{\tau
\sqrt{w_{i}}} u_{i} \hspace{0.4in}  \forall \;\; i=1, \cdots ,n, \\
%-----
y_{it} & = \left\{
\begin{array}{cc}
1 & \textrm{if} z_{it}>0, \\
0 & \text{otherwise,}
\end{array}
\right. \hspace{1.07in} \forall \;\; i=1, \cdots,n,; \; t=1,\cdots,T_{i}, \\
%-----
\alpha _{i} & \sim N\left( \overline{m}_{i}^{\prime} \zeta,
\sigma_{\alpha }^{2}\right), \hspace{0.25in} w_{it} \sim \mathcal{E}(1),  \hspace{0.75in}  u_{it} \sim N\left( 0,1\right),   \\
%-----
\beta & \sim N_{k} \left( \beta _{0}, B_{0}\right),  \hspace{0.35in} \sigma _{\alpha }^{2} \sim IG\left( \frac{c_{1}}{2}, \frac{d_{1}}{2}\right), \hspace{0.3in} \zeta \sim N_{k-1}\left( \zeta_{0}, C_{0}\right) ,
\end{split}
\label{eq:model2}
\end{equation}
%----------------------------------
where the first three lines represent the model and distribution of the mixture variables, and the last line specifies the prior distribution of the model parameters. All the notations are as described in the paper. Using the Bayes's theorem, the complete posterior density is given by,
%---------------------------------
\begin{equation}
\begin{split}
& \pi( \beta ,\alpha ,z,w,\zeta ,\sigma_{\alpha }^{2} \mid y)
\propto \bigg\{
\tprod\limits_{i=1}^{n}\tprod\limits_{t=1}^{T_{i}} \Big[ I(
z_{it}>0) I( y_{it}=1) + I( z_{it} \leq 0) I(y_{it}=0) \Big] \bigg\}  \\
%-----
& \quad \times\exp
\bigg[ -\frac{1}{2} \tsum\limits_{i=1}^{n} \Big\{ ( z_{i}-X_{i} \beta
-\iota _{T_{i}}\alpha _{i} - w_{i} \theta )^{\prime} D_{\tau \sqrt{w_{i}}}^{-2}
( z_{i}-X_{i}\beta -\iota _{T_{i}}\alpha _{i}- w_{i} \theta )
\Big\} \bigg] \\
%-----
& \quad \times \exp \bigg(
-\tsum\limits_{i=1}^{n}\tsum\limits_{t=1}^{T_{i}}w_{it}\bigg) \big( 2\pi
\sigma _{\alpha }^{2}\big)^{-\frac{n}{2}} \exp \bigg[ -\frac{1}{2\sigma
_{\alpha }^{2}}\tsum\limits_{i=1}^{n} ( \alpha _{i}-\overline{m}%
_{i}^{\prime} \zeta)^{\prime} ( \alpha _{i}-\overline{m}_{i}^{\prime}
\zeta) \bigg]  \\
%-----
& \quad \times \left( 2\pi \right) ^{-\frac{k}{2}}\left\vert B_{0}\right\vert ^{-%
\frac{1}{2}}\exp \left[ -\frac{1}{2}\left( \beta -\beta _{0}\right) ^{\prime
}B_{0}^{-1}\left( \beta -\beta _{0}\right) \right]
\left( 2\pi \right) ^{-\frac{k-1}{2}}\left\vert C_{0}\right\vert^{-%
\frac{1}{2}} \\
%-----
& \quad \times \exp \left[ -\frac{1}{2}\left( \zeta -\zeta_{0}\right)
^{\prime }C_{0}^{-1}\left( \zeta - \zeta_{0}\right) \right] \times
\left( \sigma _{\alpha }^{2}\right) ^{-\left( \frac{c_{1}}{2}+1\right) }\exp %
\left[ -\frac{d_{1}}{2\sigma _{\alpha }^{2}}\right].
\end{split}
\label{eq2}
\end{equation}%
%---------------------------------

%-------------------------------------------------------------------------------------------
\subsection{Conditional posterior density of $\beta$}
\label{subsec:beta1}
The conditional posterior density $\pi \left( \beta \mid \alpha,z, w \right)$ can be derived from the complete posterior density given by equation~\eqref{eq2} by collecting terms involving $\beta$ as follows,
%---------------------------------
\begin{equation*}
\begin{split}
\pi \left( \beta \mid \alpha,z, w \right) & \propto \exp \left[ -\frac{1}{2}
\tsum\limits_{i=1}^{n} \left( z_{i}-X_{i}\beta -\iota _{T_{i}}\alpha _{i} - w_{i} \theta \right)^{\prime }D_{\tau \sqrt{w_{i}}%
}^{-2} \left( z_{i}-X_{i}\beta -\iota _{T_{i}}\alpha _{i}- w_{i} \theta \right) \right]  \\
& \quad \times \exp \left[ -\frac{1}{2}\left( \beta -\beta _{0}\right) ^{\prime}B_{0}^{-1}\left( \beta -\beta _{0}\right) \right].
\end{split}
\end{equation*}
%---------------------------------
Let $G_i = \left( z_{i}-\iota _{T_{i}}\alpha _{i} - w_{i} \theta \right)$, and consider the expressions in the exponential without the $-1/2$ term. Then the quadratic expressions can be opened and regrouped as,
%---------------------------------
\begin{equation*}
\begin{split}
& \tsum\limits_{i=1}^{n} \left(G_i - X_{i}\beta \right)^{\prime } D_{\tau \sqrt{w_{i}}}^{-2} \left(G_i - X_{i}\beta \right) + \left( \beta -\beta _{0}\right)^{\prime}B_{0}^{-1}\left( \beta -\beta _{0}\right)\\
%----
& =  \tsum\limits_{i=1}^{n} G^{\prime}_i  D_{\tau \sqrt{w_{i}}}^{-2} G_i - 2 \tsum\limits_{i=1}^{n} \beta^{\prime} X^{\prime}_{i}  D_{\tau \sqrt{w_{i}}}^{-2} G_i +
     \tsum\limits_{i=1}^{n} \beta^{\prime} X^{\prime}_{i}  D_{\tau \sqrt{w_{i}}}^{-2} X_{i}\beta + \beta^{\prime}B_{0}^{-1} \beta - 2  \beta^{\prime}B_{0}^{-1} \beta _{0} + \beta^{\prime} _{0}B_{0}^{-1} \beta _{0} \\
%-----
& = \beta^{\prime}  \left(  \tsum\limits_{i=1}^{n}X^{\prime}_{i}  D_{\tau \sqrt{w_{i}}}^{-2} X_{i} + B_{0}^{-1} \right) \beta  -2  \beta^{\prime}  \left(  \tsum\limits_{i=1}^{n}X^{\prime}_{i}  D_{\tau \sqrt{w_{i}}}^{-2} G_{i} +  B_{0}^{-1}  \beta _{0} \right)
     +  \tsum\limits_{i=1}^{n} G^{\prime}_i  D_{\tau \sqrt{w_{i}}}^{-2} G_i + \beta^{\prime} _{0}B_{0}^{-1} \beta _{0}.
\end{split}
\end{equation*}
%---------------------------------
Our interest lies in the distribution of $\beta$, so all terms that do not involve $\beta$ are absorbed into the proportionality constant. Applying the idea of completing the square to the previous equation we have,
%---------------------------------
\begin{equation*}
\begin{split}
& \pi \left( \beta \mid \alpha,z, w \right)  \propto \exp \left[ -\frac{1}{2} \left( \beta -  \widetilde{\beta} \right)^{\prime } \widetilde{B}^{-1}  \left( \beta -  \widetilde{\beta} \right) \right], \quad \textrm{where}, \\
& \widetilde{B}^{-1}= \left(  \tsum\limits_{i=1}^{n} X^{\prime}_{i} D_{\tau \sqrt{w_{i}}}^{-2}  X_{i}  + B^{-1}_{0} \right), \quad \textrm{and} \quad \widetilde{\beta}= \widetilde{B} \left(  \tsum\limits_{i=1}^{n} X^{\prime}_{i} D_{\tau \sqrt{w_{i}}}^{-2}
\left( z_{i} - \iota _{T_{i}}\alpha _{i} - w_{i}    \theta  \right)  + B^{-1}_{0} \beta_0 \right),
\end{split}
\end{equation*}
%---------------------------------
which is recognized as the kernel of a normal distribution. Hence we can write,
%---------------------------------
\begin{equation}
\begin{array}{c}
\beta \mid \alpha, z, w  \sim N_{k}\left( \widetilde{\beta}, \widetilde{B}\right), \vspace{5px}\\
\textrm{where} \;\; \widetilde{B}^{-1}= \left(  \tsum\limits_{i=1}^{n} X^{\prime}_{i} D_{\tau \sqrt{w_{i}}}^{-2}  X_{i}  + B^{-1}_{0} \right), \;\; \text{and} \;\; \widetilde{\beta}= \widetilde{B} \left(  \tsum\limits_{i=1}^{n} X^{\prime}_{i} D_{\tau \sqrt{w_{i}}}^{-2}
 \left( z_{i} - \iota _{T_{i}}\alpha _{i} - w_{i}    \theta  \right)  + B^{-1}_{0} \beta_0 \right).
\end{array}
\label{eq3}
\end{equation}.
%---------------------------------

%-----------------------------------------------------------------------------------------
\subsection{Conditional posterior density of $\alpha_i$}
\label{subsec:alpha1}

Similar to $\beta$, the conditional posterior density $\pi \left( \alpha_i \mid \beta, z, w, \sigma _{\alpha }^{2}, \zeta \right)$ can also be derived from the complete posterior density, given by equation~\eqref{eq2}, by collecting terms involving $\alpha_{i}$. This results in the following expression,
%---------------------------------
\begin{equation*}
\begin{split}
\pi \left( \alpha_i \mid \beta, z, w, \sigma _{\alpha }^{2}, \zeta\right) & \propto \exp \left[ -\frac{1}{2}  \left( z_{i}-X_{i}\beta -\iota _{T_{i}}\alpha _{i} - w_{i} \theta \right)^{\prime }D_{\tau \sqrt{w_{i}}%
}^{-2} \left( z_{i}-X_{i}\beta -\iota _{T_{i}}\alpha _{i}- w_{i} \theta \right) \right]  \\
%----
& \quad \times \exp \left[ -\frac{1}{2\sigma_{\alpha }^{2}} \left( \alpha _{i}-\overline{m}_{i}^{\prime }\zeta \right) ^{\prime }\left( \alpha _{i}-\overline{m}_{i}^{\prime }\zeta \right) \right]
\end{split}
\end{equation*}
%---------------------------------
Let $H_i = \left( z_{i} - X_{i}\beta  - w_{i} \theta \right)$ and consider the expressions in the exponential without the constant $-1/2$. Then the expression can be simplified as,
%---------------------------------
\begin{equation*}
\begin{split}
& \left(H_i - \iota _{T_{i}}\alpha _{i} \right)^{\prime } D_{\tau \sqrt{w_{i}}}^{-2} \left(H_i - \iota _{T_{i}}\alpha _{i} \right) +  \sigma_{\alpha }^{-2} \left( \alpha _{i}-\overline{m}_{i}^{\prime }\zeta \right) ^{\prime }\left( \alpha _{i}-\overline{m}_{i}^{\prime }\zeta \right) \\
%-----
& =  H^{\prime}_i  D_{\tau \sqrt{w_{i}}}^{-2} H_i - 2 \alpha^{\prime}_{i} \iota^{\prime}_{T_{i}} D_{\tau \sqrt{w_{i}}}^{-2} H_i +   \alpha^{\prime}_{i} \iota^{\prime}_{T_{i}} D_{\tau \sqrt{w_{i}}}^{-2}  \iota _{T_{i}}\alpha _{i}
 +  \sigma_{\alpha }^{-2} \alpha^{\prime}_{i} \alpha_{i} - 2 \sigma_{\alpha }^{-2} \alpha^{\prime}_{i} \overline{m}_{i}^{\prime }\zeta +  \sigma_{\alpha }^{-2}  \zeta^{\prime} \overline{m}_{i} \overline{x}_{i}^{\prime }\zeta \\
%-----
& = \alpha^{\prime}_{i} \left(   \iota^{\prime}_{T_{i}} D_{\tau \sqrt{w_{i}}}^{-2}  \iota _{T_{i}} +   \sigma_{\alpha }^{-2}  \right)\alpha _{i} - 2  \alpha^{\prime}_{i} \left(   \iota^{\prime}_{T_{i}} D_{\tau \sqrt{w_{i}}}^{-2}  \iota _{T_{i}} +  \sigma_{\alpha }^{-2}  \overline{m}_{i}^{\prime }\zeta \right) + H^{\prime}_i  D_{\tau \sqrt{w_{i}}}^{-2} H_i + \sigma_{\alpha }^{-2}  \zeta^{\prime} \overline{m}_{i} \overline{m}_{i}^{\prime }\zeta.
\end{split}
\end{equation*}
%---------------------------------
We are interested in the distribution of $\alpha_i$, so all terms that do not involve $\alpha_i$ are absorbed into the proportionality constant. Applying the idea of completing the square as in the previous equation we have,
%---------------------------------
\begin{equation*}
\begin{split}
& \pi \left( \alpha_i \mid \beta, z, w, \sigma _{\alpha }^{2}, \zeta \right)  \propto \exp \left[ -\frac{1}{2} \left( \alpha_i-  \widetilde{a} \right)^{\prime } \widetilde{A}^{-1}  \left( \alpha_i -  \widetilde{a} \right) \right], \quad \textrm{where}, \\
& \widetilde{A}^{-1}= \left(   \iota^{\prime}_{T_{i}} D_{\tau \sqrt{w_{i}}}^{-2}  \iota _{T_{i}} +   \sigma_{\alpha }^{-2} \right), \;\; \text{and} \;\; \widetilde{a}= \widetilde{A} \left(  \iota^{\prime}_{T_{i}} D_{\tau \sqrt{w_{i}}}^{-2}  \left( z_{i} - X_{i}\beta  - w_{i} \theta \right)+   \sigma_{\alpha }^{-2} \overline{m}_{i}^{\prime }\zeta \right),
\end{split}
\end{equation*}
%---------------------------------
which is the kernel of a normal distribution. Hence, the conditional posterior density of $\alpha_{i}$ is given by,
%--------------------------------
\begin{equation}
\begin{array}{c}
 \alpha_i \mid \beta, z, w, \sigma _{\alpha }^{2}, \zeta \sim N\left( \widetilde{a}, \widetilde{A}\right), \vspace{5px} \\
\textrm{where} \;\; \widetilde{A}^{-1}= \left(\iota^{\prime}_{T_{i}} D_{\tau \sqrt{w_{i}}}^{-2}  \iota _{T_{i}} +   \sigma_{\alpha }^{-2} \right), \;\; \text{and} \;\; \widetilde{a} = \widetilde{A} \left(  \iota^{\prime}_{T_{i}} D_{\tau \sqrt{w_{i}}}^{-2}  \left( z_{i} - X_{i}\beta  - w_{i} \theta \right) + \sigma_{\alpha }^{-2} \overline{m}_{i}^{\prime }\zeta \right).
\end{array}
\label{eq4}
\end{equation}
%--------------------------------

%-----------------------------------------------------------------------------------------------

\subsection{Conditional posterior density of $w_{it}$}
\label{subsec:w1}

The conditional posterior density $\pi \left( w_{it} \mid \beta, \alpha_i , z_{it} \right)$ can be obtained from the complete posterior density (equation~\ref{eq2}) by collecting terms involving $w_{it}$. This is done as follows,
%---------------------------------
\begin{equation*}
\begin{split}
\pi \left( w_{it} \mid  \beta, \alpha_i , z_{it} \right) & \propto  w^{-1/2}_{it} \exp \left[ -\frac{1}{2}  \left( z_{it} - x^{\prime}_{it}\beta - \alpha _{i} - w_{it} \theta \right)^{\prime } \frac{1}{\tau^2 w_{it}} \left( z_{it} - x^{\prime}_{it}\beta - \alpha _{i} - w_{it} \theta \right) \right]  \times \exp \left[ - w_{it} \right] \\
%----
&  \propto w^{-1/2}_{it} \exp \left[ -\frac{1}{2} \left\{ 2  w_{it} +  \frac{ \left( z_{it} - x^{\prime}_{it}\beta - \alpha _{i} - w_{it} \theta \right)^{2}}{\tau^2 w_{it}} \right\}  \right] \\
%----
&  \propto w^{-1/2}_{it} \exp \left[ -\frac{1}{2} \left\{ \left(   \frac{\theta^2}{\tau^2} + 2\right) w_{it} +  \frac{1}{\tau^2} \frac{   \left( z_{it} - x^{\prime}_{it}\beta - \alpha _{i} \right)^{2}  }{w_{it}} \right\}  \right],
\end{split}
\end{equation*}
%---------------------------------
where all terms not involving $w_{it}$ have been absorbed in the proportionality constant. By comparing the last expression to that of a generalized inverse Gaussian distribution ($GIG$)\footnote{The generalized inverse Gaussian distribution $GIG(p,a,b)$ is a three-parameters family of continuous probability distributions with probability density,
%---------------------------------
\begin{equation*}
f(x) = \frac{(b/a)^{p/2}}{2 K_p \left( \sqrt{ab} \right) } x^{(p-1)} \exp \left[ -\frac{1}{2}  \left( \frac{a}{x} + bx \right) \right],  \qquad x > 0,
\end{equation*}
%---------------------------------
where $K_p \left(. \right)$ is the modified Bessel function of the second kind, $a>0$, $b>0$ and $p$ is a real parameter.}, we have
%---------------------------------
\begin{equation}
\begin{split}
w_{it} \mid \beta, \alpha_i , z_{it}  & \sim GIG\left( \frac{1}{2}, \widetilde{\lambda}_{it}, \widetilde{\eta} \right) \quad \textrm{for} \; i=1,...,n, \; \textrm{and} \; t=1,...,T_i,\\
\text{where} \;\; \widetilde{\lambda}_{it} & = \left(  \frac{   z_{it}-x^{\prime}_{it} \beta  -\alpha_i}{\tau} \right)^2, \;\; \text{and} \;\; \widetilde{\eta}= \left( \frac{\theta^2}{\tau^2} + 2 \right).
\end{split}
\label{eq5}
\end{equation}
%---------------------------------

%--------------------------------------------------------------------------------------------

\subsection{Conditional posterior density of $\sigma_{\alpha }^{2}$}
\label{subsec:sigma1}
The conditional posterior density $\pi \left( \sigma_{\alpha }^{2} \mid \alpha, \zeta  \right)$ is also derived from the complete posterior density (equation~\ref{eq2}) by collecting terms involving $\sigma_{\alpha}^{2}$. This is done as follows,
%---------------------------------
\begin{equation*}
\begin{split}
\pi \left( \sigma _{\alpha }^{2} \mid \alpha, \zeta  \right) & \propto \left( 2\pi \sigma _{\alpha }^{2}\right) ^{-\frac{n}{2}} \exp \left[ -\frac{1}{2\sigma
_{\alpha }^{2}}\tsum\limits_{i=1}^{n}\left( \alpha _{i}-\overline{m}_{i}^{\prime }\zeta \right) ^{\prime }\left( \alpha _{i}-\overline{m}_{i}^{\prime }\zeta \right) \right]
%----
\times \left( \sigma _{\alpha }^{2}\right) ^{-\left( \frac{c_{1}}{2}+1\right) }\exp \left[ -\frac{d_{1}}{2\sigma _{\alpha }^{2}}\right]   \\
%-----
& \propto \left( 2\pi \sigma _{\alpha }^{2}\right)^{-\frac{(n+c_1)}{2} + 1} \exp \left[ -\frac{1}{2\sigma
_{\alpha }^{2}} \left\{ d_1 + \tsum\limits_{i=1}^{n}\left( \alpha _{i}-\overline{m}_{i}^{\prime }\zeta \right) ^{\prime }\left( \alpha _{i}-\overline{m}_{i}^{\prime }\zeta \right) \right\} \right],
\end{split}
\end{equation*}
%---------------------------------
which is the kernel of an inverse-gamma distribution ($IG$) with shape parameter $\widetilde{c}_1 = n+c_1 $ and scale parameter $\widetilde{d}_1 = d_1 +  \tsum\limits_{i=1}^{n}\left( \alpha _{i}-\overline{m}_{i}^{\prime }\zeta \right) ^{\prime }\left( \alpha _{i}-\overline{m}_{i}^{\prime }\zeta \right)$. Therefore, the conditional posterior density can be written as,
%---------------------------------
\begin{equation}
\begin{array}{c}
\sigma _{\alpha }^{2} \mid \alpha, \zeta   \sim IG\left( \frac{ \widetilde{c}_1}{2}, \frac{ \widetilde{d}_1}{2} \right),\\
  \text{where} \;\;  \widetilde{c}_1 = (n + c_1 ), \;\; \text{and} \;\; \widetilde{d}_1= d_1 + \tsum\limits_{i=1}^{n}  \left( \alpha _{i}-\overline{m}_{i}^{\prime }\zeta \right)^{\prime }\left( \alpha _{i}-\overline{m}_{i}^{\prime }\zeta \right).
\end{array}
\label{eq6}
\end{equation}
%---------------------------------

%----------------------------------------------------------------------------------------------
\subsection{Conditional posterior density of $\zeta$}
\label{subsec:zeta1}
The conditional posterior density $\pi \left( \zeta \mid \alpha, \sigma_{\alpha }^{2} \right)$ is derived from the complete posterior density (equation~\ref{eq2}). Collecting terms involving $\zeta$ we obtain the following expression,
%---------------------------------
\begin{equation*}
\pi \left(  \zeta \mid \alpha, \sigma_{\alpha}^{2} \right) \propto  \exp \left[ -\frac{1}{2\sigma_{\alpha }^{2}} \tsum\limits_{i=1}^{n} \left( \alpha _{i}-\overline{m}_{i}^{\prime }\zeta \right) ^{\prime }\left( \alpha _{i}-\overline{m}_{i}^{\prime }\zeta \right) \right] \\
\times \exp \left[ -\frac{1}{2}\left( \zeta -\zeta_{0} \right)^{\prime} C_{0}^{-1} \left( \zeta -\zeta_{0}\right). \right]
\end{equation*}
%---------------------------------
Focusing on the expressions in the exponential but without the $-1/2$ term, we have,
%---------------------------------
\begin{equation*}
\begin{split}
& \sigma_{\alpha }^{-2} \tsum\limits_{i=1}^{n} \left( \alpha _{i}-\overline{m}_{i}^{\prime }\zeta \right) ^{\prime }\left( \alpha _{i}-\overline{m}_{i}^{\prime }\zeta \right) + \left( \zeta - \zeta_{0} \right)^{\prime} C_{0}^{-1}\left( \zeta - \zeta_{0} \right) \\
%----
& =  \sigma_{\alpha }^{-2} \tsum\limits_{i=1}^{n} \alpha^{\prime}_{i} \alpha _{i} - 2  \sigma_{\alpha }^{-2} \tsum\limits_{i=1}^{n} \zeta^{\prime} \overline{m}_{i} \alpha _{i}
    +  \sigma_{\alpha }^{-2} \tsum\limits_{i=1}^{n} \zeta^{\prime} \overline{m}_{i} \overline{m}_{i}^{\prime }\zeta + \zeta^{\prime}C_{0}^{-1} \zeta - 2 \zeta^{\prime}C_{0}^{-1} \zeta_{0} + \zeta_{0}^{\prime}C_{0}^{-1} \zeta_{0 } \\
%-----
& =  \zeta^{\prime} \left(  \sigma_{\alpha }^{-2} \tsum\limits_{i=1}^{n}  \overline{m}_{i} \overline{m}_{i}^{\prime } +   C_{0}^{-1}   \right) \zeta -2  \zeta^{\prime} \left(  \sigma_{\alpha }^{-2} \tsum\limits_{i=1}^{n}  \overline{m}_{i} \alpha _{i} + C_{0}^{-1} \zeta_{0} \right) +   \sigma_{\alpha }^{-2} \tsum\limits_{i=1}^{n} \alpha^{\prime} _{i} \alpha _{i}  + \zeta_{0}^{\prime}C_{0}^{-1} \zeta_{0}
\end{split}
\end{equation*}
%---------------------------------

The interest lies in the distribution of $\zeta$, so all terms that do not involve $\zeta$ are absorbed into the proportionality constant. Applying the idea of completing the square we have,
%---------------------------------
\begin{equation*}
\begin{split}
& \pi \left( \zeta \mid \alpha, \sigma_{\alpha }^{2}  \right)  \propto \exp \left[ -\frac{1}{2} \left( \zeta -  \widetilde{\zeta} \right)^{\prime } \widetilde{\Sigma}_{\zeta}^{-1}  \left( \zeta -  \widetilde{\zeta} \right) \right], \qquad \textrm{where} \\
& \widetilde{\Sigma}_{\zeta}^{-1} = \left(  \sigma_{\alpha }^{-2} \tsum\limits_{i=1}^{n}  \overline{m}_{i} \overline{m}_{i}^{\prime } +   C_{0}^{-1} \right), \quad \text{and} \quad \widetilde{\zeta}=  \widetilde{\Sigma}_{\zeta} \left(   \sigma _{\alpha }^{-2}  \tsum\limits_{i=1}^{n} \overline{m}_{i} \alpha^{\prime}_{i} +  C^{-1}_{0} \zeta_{0 } \right),
\end{split}
\end{equation*}
%---------------------------------
which is recognized as the kernel of a normal distribution. Hence, the conditional posterior density is,
%---------------------------------
\begin{equation}
\begin{array}{c}
\zeta \mid \alpha, \sigma_{\alpha }^{2} \sim N_{k-1}\left( \widetilde{\zeta}, \widetilde{\Sigma}_{\zeta}\right),\\
  \text{where}  \;\;  \widetilde{\Sigma}^{-1}_{\zeta}  = \left(   \sigma _{\alpha }^{-2}  \tsum\limits_{i=1}^{n} \overline{m}_{i} \overline{m}_{i}^{\prime } +  C^{-1}_{0} \right),
  \;\; \text{and} \;\; \widetilde{\zeta} = \widetilde{\Sigma}_{\zeta} \left( \sigma _{\alpha }^{-2}  \tsum\limits_{i=1}^{n} \overline{m}_{i} \alpha^{\prime}_{i} +  C^{-1}_{0} \zeta_{0}\right).
\end{array}
\label{eq7}
\end{equation}
%---------------------------------

%-------------------------------------------------------------------------------------------------
\subsection{Conditional posterior density of $z_{it}$}
\label{subsec:z1}
The conditional posterior density $\pi \left( z_{it} \mid  \beta, \alpha, w , y_{it}\right)$ is obtained by collecting terms involving $z_{it}$ from the complete posterior density (equation~\ref{eq2}) and seen to have the following expression,
%---------------------------------
\begin{equation*}
\begin{split}
\pi \left( z_{it} \mid  \beta, \alpha, w, y_{it} \right) & \propto  \left[ I\left( z_{it}>0\right) I\left( y_{it}=1\right) +I\left( z_{it}\leq 0\right) I\left( y_{it}=0\right) \right] \\
& \quad \times \exp \left[ -\frac{1}{2}  \left( z_{it} - x^{\prime}_{it}\beta - \alpha _{i} - w_{it} \theta \right)^{\prime } \frac{1}{\tau^2 w_{it}} \left( z_{it} - x^{\prime}_{it}\beta - \alpha _{i} - w_{it} \theta \right) \right]  \\
%----
& \propto \left[ I\left( z_{it}>0\right) I\left( y_{it}=1\right) +I\left( z_{it}\leq 0\right) I\left( y_{it}=0\right) \right]  \phi \left( z_{it} | x^{\prime}_{it}\beta + \alpha _{i} + w_{it} \theta,  \tau^2 w_{it}  \right),
\end{split}
\end{equation*}
%---------------------------------
where the notation  $\phi \left( x| \mu, \Sigma \right)$ denotes the probability density function of a normal distribution with mean $\mu$ and variance $\Sigma$. Note that the observations are conditionally independent, so we can draw each $z_{it}$ independently of each other. Consequently, we have,
%---------------------------------
\begin{equation*}
\begin{split}
\pi \left( z_{it} \mid  \beta, \alpha, w, y_{it} \right) & \propto    I\left( z_{it} \leq 0\right)\phi \left( z_{it}| x^{\prime}_{it}\beta + \alpha _{i} + w_{it} \theta,  \tau^2 w_{it}  \right) \quad \text{if} \; y_{it}=0, \\
%-----
\pi \left( z_{it} \mid  \beta, \alpha, w, y_{it}  \right) & \propto    I\left( z_{it}>0\right)\phi \left( z_{it}|  x^{\prime}_{it}\beta + \alpha _{i} + w_{it} \theta,  \tau^2 w_{it}  \right) \quad \text{if} \; y_{it}=1,
\end{split}
\end{equation*}
%---------------------------------
which implies that we can draw $z_{it}$ as follows,
%---------------------------------
\begin{equation}
z_{it} \mid  \beta, \alpha, w, y_{it}  \sim  \left\{
\begin{array}{cc}
TN_{\left(- \infty , 0 \right] } \left( x^{\prime}_{it} \beta  + \alpha_i  + w_{it} \theta , \tau^2 w_{it} \right) & \text{ if } y_{it}=0, \vspace{5px} \\
TN_{\left( 0, \infty \right) } \left( x^{\prime}_{it} \beta  + \alpha_i  + w_{it} \theta , \tau^2 w_{it} \right) & \text{ if }y_{it}=1,
\end{array}%
\right.
\label{eq8}
\end{equation}
%---------------------------------
where $TN_{\left[a,b \right] } \left( \mu, \sigma^2 \right)$ denotes a normal distribution truncated to the interval $\left[a, b\right]$ with mean $\mu$ and variance $\sigma^2$.

%-------------------------------------------------------------------------------------------------
%                                 Blocked Sampling Algorithm
%-------------------------------------------------------------------------------------------------
\section{Conditional densities for the blocked sampling in the BPQRCRE model}
\label{sec:blocked}

\subsection{Joint posterior conditional density of $(\beta , z_{it} )$}
\label{subsec:joint1}
The Gibbs sampler presented in Section~\ref{sec:nonblocked} is straightforward, however, there is a potential for poor mixing due to correlation between ($\beta $, $\alpha _{i}$) and ($z_{i}$, $\alpha _{i}$). This correlation arises because the variables corresponding to the parameters in $\alpha _{i}$ are often a subset of those in $x_{it}^{\prime }$. Thus, by conditioning these items on one another, the mixing of the Markov chain will be slow.

To avoid this issue, we present an alternative algorithm which jointly samples ($\beta $, $z_{i}$) in one block within the Gibbs sampler. In particular, $\beta$ is sampled marginally of $\alpha_i$ from a multivariate normal distribution. Thereafter, the latent variable $z_i$ is sampled marginally of $\alpha_i$ from a truncated multivariate normal distribution.

\subsubsection{Conditional posterior density of $\beta$} \label{subsec:beta2}

First, we derive the conditional posterior density of $\beta$, marginalized over the random effects $\alpha_i$. Since $\alpha _{i}\sim N\left( \overline{m}_{i}^{\prime }\zeta ,\sigma
_{\alpha }^{2}\right)$, we can write $\alpha _{i}=  \overline{m}_{i}^{\prime }\zeta + \xi_i$ with $\xi_{i}\sim N\left(0,\sigma_{\alpha }^{2}\right)$. Therefore, the model can be rewritten as follows,
%--------------------------------
\begin{equation*}
\begin{split}
z_{i} & =  X_{i}\beta +\iota _{T_{i}}\alpha _{i}+ w_{i} \theta+D_{\tau \sqrt{w_{i}}}u_{i},  \qquad \forall \; i=1,...,n, \\
 & =  X_{i} \beta +\iota _{T_{i}} \overline{m}_{i}^{\prime }\zeta +  w_{i} \theta + v_i,
\end{split}
\end{equation*}%
%--------------------------------
where $v_i= \iota _{T_{i}} \xi_i + D_{\tau \sqrt{w_{i}}}u_{i}$. Given the  properties of the components that constitutes $\nu_{i}$, we have $E(\nu_{i})=0_{T_{i}}$. The covariance can be derived to have the following expression,
%--------------------------------
\begin{equation*}
\begin{split}
Cov(\nu_{i}) & = E \left[ v_i v^{\prime}_i \right]  = E \left[ \iota _{T_{i}} \xi_i \xi^{\prime}_i  \iota^{\prime}_{T_{i}} + D_{\tau \sqrt{w_{i}}}u_{i} u^{\prime}_{i} D_{\tau \sqrt{w_{i}}} \right] \\
& = \sigma _{\alpha }^{2}  \iota _{T_{i}} \iota^{\prime} _{T_{i}}
    + D_{\tau \sqrt{w_{i}}}^{-2}
=  \sigma _{\alpha }^{2}  J_{T_{i}} + D_{\tau \sqrt{w_{i}}}^{-2}=\Omega_i.
\end{split}
\end{equation*}%
%--------------------------------
where $J_{T_{i}} = \iota _{T_{i}} \iota^{\prime} _{T_{i}}$. The conditional posterior density $\pi \left( \beta \mid z, w, \sigma _{\alpha }^{2}, \zeta \right)$, marginally of $\alpha$, has the expression,
%--------------------------------
\begin{equation*}
\begin{split}
\pi \left( \beta \mid z, w, \sigma _{\alpha }^{2} \right) & \propto  \exp \left[ -\frac{1}{2} \tsum\limits_{i=1}^{n} \left( z_{i}-X_{i}\beta -\iota _{T_{i}} \overline{m}_{i}^{\prime }\zeta - w_{i} \theta \right)^{\prime } \Omega_{i}^{-1} \left( z_{i}-X_{i}\beta
-\iota _{T_{i}}\overline{m}_{i}^{\prime }\zeta- w_{i} \theta \right) \right]  \\
%----
&  \quad \times \exp \left[ -\frac{1}{2}\left( \beta -\beta _{0}\right) ^{\prime}B_{0}^{-1}\left( \beta -\beta _{0}\right) \right].
\end{split}
\end{equation*}
%--------------------------------
Similar to Section~\ref{subsec:beta1}, we open the quadratic expressions and collect the terms involving $\beta$ to complete a square. This yields the following,
%--------------------------------
\begin{equation*}
\begin{array}{c}
\pi \left( \beta \mid z, w, \sigma _{\alpha }^{2}, \zeta \right)  \propto \exp \left[ -\frac{1}{2} \left( \beta -  \widetilde{\beta} \right)^{\prime } \widetilde{B}^{-1}  \left( \beta -  \widetilde{\beta} \right) \right] \vspace{5px} \\
%-----
\text{where} \;\; \widetilde{B}^{-1}= \left(  \tsum\limits_{i=1}^{n} X^{\prime}_{i} \Omega^{-1}_{i}  X_{i}  + B^{-1}_{0} \right),  \;\; \text{and} \;\; \widetilde{\beta}=  \widetilde{B} \left(  \tsum\limits_{i=1}^{n} X^{\prime}_{i} \Omega^{-1}_{i} \left( z_{i} - \iota _{T_{i}}  \overline{m}_{i}^{\prime } \zeta -  w_{i} \theta  \right)  + B^{-1}_{0} \beta_0 \right),
\end{array}
\end{equation*}
%--------------------------------
which is clearly the kernel of a normal distribution. Hence, the conditional posterior density is,
%--------------------------------
\begin{equation}
\begin{array}{c}
\beta \mid z, w, \sigma _{\alpha }^{2}, \zeta \sim N_{k}\left( \widetilde{\beta}, \widetilde{B}\right) \vspace{5px} \\
\text{where} \;\;  \widetilde{B}^{-1} = \left(  \tsum\limits_{i=1}^{n} X^{\prime}_{i} \Omega^{-1}_{i}  X_{i}  + B^{-1}_{0} \right), \;\; \text{and} \;\; \widetilde{\beta}=  \widetilde{B} \left(  \tsum\limits_{i=1}^{n} X^{\prime}_{i} \Omega^{-1}_{i} \left( z_{i} - \iota _{T_{i}}  \overline{m}_{i}^{\prime } \zeta -  w_{i}  \theta  \right)  + B^{-1}_{0} \beta_0 \right).
\end{array}
\label{eq9}
\end{equation}
%--------------------------------

%-----------------------------------------------------------------------------

\subsubsection{Conditional posterior density of $z_i$} \label{subsec:z2}

The conditional posterior density of the latent variable $z$, marginally of $\alpha$, has the following expression,
%--------------------------------
\begin{equation*}
\begin{split}
\pi \left( z \mid \beta, z, w, \zeta, \sigma_{\alpha }^{2}, y\right)  &
\propto
\tprod\limits_{i=1}^{n}  \left\{   \tprod\limits_{t=1}^{T_{i}}\left[ I\left(
z_{it}>0\right) I\left( y_{it}=1\right) +I\left( z_{it}\leq 0\right) I\left(
y_{it}=0\right) \right]
\right\}  \\
%-----
& \times  \exp \left[ -\frac{1}{2}  \left( z_{i}-X_{i}\beta -\iota _{T_{i}} \overline{m}_{i}^{\prime }\zeta - w_{i} \theta \right)^{\prime } \Omega_{i}^{-1} \left( z_{i}-X_{i}\beta
-\iota _{T_{i}}\overline{m}_{i}^{\prime }\zeta- w_{i} \theta \right) \right].
\end{split}
\end{equation*}%
%--------------------------------
The above expression is recognized as the kernel of a truncated multivariate normal (TMVN) distribution. Hence, we can write,
%--------------------------------
\begin{equation*}
z_{i} \mid  \beta, w_i, \zeta, \sigma_{\alpha }^{2}, y_{i}  \sim TMVN_{B_i} \left( X_i \beta +  \iota _{T_{i}}    \overline{m}_{i}^{\prime } \zeta + w_{i} \theta , \Omega_i   \right) \qquad \forall \;\; i=1,\cdots,n,
\end{equation*}
%--------------------------------
where $B_i=\left( B_{i1} \times  B_{i2} \times ... \times B_{iT_i} \right)$ and $B_{it}$ is the interval $\left(0, \infty \right)$ if $y_{it}=1$ and the interval $\left(-\infty, 0 \right]$ if $y_{it}=0$. Sampling directly from a $TMVN$ is not possible, hence, as in \cite{Rahman-Vossmeyer-2019}, we resort to the method proposed in \cite{geweke1991,geweke2005}, which utilizes Gibbs sampling to make draws from a $TMVN$.

We apply the theorem on the conditional multivariate normal distribution (see \cite{geweke2005} p.171)\footnote{Let $\begin{pmatrix}x\\y\end{pmatrix} \sim N\left(\mu , \Sigma \right)$ with $\mu = \begin{pmatrix}\mu_x\\ \mu_y\end{pmatrix}$ and $\Sigma = \begin{bmatrix}\Sigma_{xx}&\Sigma_{xy}\\\Sigma_{yx}&\Sigma_{yy}\end{bmatrix}$, then the distribution of $y$ conditional on $x$ is normal $N\left( \mu_{y.x}  , \Sigma_{y.x}  \right)$ with $\mu_{y.x} = \mu_y + \Sigma_{yx} \Sigma_{xx}^{-1} \left(x - \mu \right)$ and $\Sigma_{y.x} = \Sigma_{yy} - \Sigma_{yx}  \Sigma_{xx}^{-1}  \Sigma_{xy}$.}
%--------------------------------
on full conditional $f \left( z_{it} \mid  z_{i,-t} \right) = f \left( z_{it} \mid  z_{i1}, ..., z_{i(t-1)},  z_{i(t+1)}, ..., z_{iT_i} \right)$ which we denote as $t \mid -t$. This results in:
%-------------------------------
\begin{equation*}
\begin{pmatrix} z_{it}\\  z_{i,-t} \end{pmatrix} \sim N\left(\mu , \Sigma\right), \;\; \textrm{where} \;\; \mu = \begin{pmatrix} x^{\prime}_{it} \beta + \overline{m}_{i}^{\prime } \zeta + w_{it}\theta \\  \left(  X_i \beta +  \iota _{T_{i}}  \overline{m}_{i}^{\prime } \zeta + w_{i} \theta  \right)_{-t} \end{pmatrix}, \;\; \textrm{and} \;\; \Sigma = \begin{bmatrix}\Sigma_{t, t}&\Sigma_{t, -t}\\\Sigma_{-t, t} & \Sigma_{-t, -t}\end{bmatrix},
\end{equation*}
%--------------------------------
where $\left(  X_i \beta +  \iota_{T_{i}}  \overline{m}_{i}^{\prime } \zeta + w_{i} \theta  \right)_{-t}$ is a column vector with $t$-th element removed. $\Sigma_{t, t}$ denotes the $(t,t)$-th element of $\Omega_i$, $\Sigma_{t, -t}$ denotes the $t$-th row of  $\Omega_i$ with element in the $t$-th column removed and $\Sigma_{-t, -t}$ is the $\Omega_i$ matrix with $t$-th row and $t$-th column removed. Then, the distribution of $z_{it}$ conditional on $z_{i,-t}$ is normal with conditional mean and conditional variance equal to,
%--------------------------------
\begin{eqnarray*}
\mu_{t \mid -t} & = & x^{\prime}_{it} \beta + \overline{m}_{i}^{\prime } \zeta + w_{it}\theta + \Sigma_{t, -t} \Sigma^{-1}_{-t , -t} \left( z_{i,-t}  - \left(  X_i \beta +  \iota _{T_{i}}    \overline{m}_{i}^{\prime } \zeta + w_{i} \theta  \right)_{-t}    \right), \\
\Sigma_{t \mid -t} & = &  \Sigma_{t, t} -  \Sigma_{t, -t} \Sigma^{-1}_{-t , -t} \Sigma_{-t, t}.
\end{eqnarray*}
%--------------------------------

We can then construct a Markov chain which continuously draws from   $f \left( z_{it} \mid  z_{i,-t} \right)$ subject to the bound $\left(0, \infty \right)$ if $y_{it}=1$ or $\left(-\infty, 0 \right]$ if $y_{it}=0$. This is done by sampling $z_i$ at the $j$-th pass of the MCMC iteration using a series of conditional posterior distributions as follows:
%--------------------------------
\begin{equation}
z^{j}_{it} \mid  z^{j}_{i1}, ..., z^{j}_{i(t-1)},z^{j}_{i(t+1)}, ..., z^{j}_{iT_i} \sim TN_{B_{it}} \left(\mu_{t \mid -t}, \Sigma_{t \mid -t} \right), \quad \mathrm{for} \;\; t=1,\cdots,T_i,
\label{eq10}
\end{equation}
%--------------------------------
where $TN$ denotes a truncated normal distribution. The terms $ \mu_{t \mid -t}$ and $ \Sigma_{t \mid -t}$ are the conditional mean and variance defined above and $z_{i,-t}$ at the $j-$th pass is $z^{j}_{i,-t}=\left( z^{j}_{i1}, ..., z^{j}_{i(t-1)},  z^{j-1}_{i(t+1)}, ..., z^{j-1}_{iT_i} \right)^{\prime}$.

%-------------------------------------------------------------------------------

\subsection{Other conditional posterior densities}
The derivations for the conditional posterior densities of $\alpha_i$, $w_{it}$, $\sigma _{\alpha }^{2}$ and $\zeta$ remains unaltered as presented earlier from Section~\ref{subsec:alpha1} to Section~\ref{subsec:zeta1}. For the sake of completion, we present the conditional distributions below once more.
%---------------------------------
\begin{equation}
\begin{array}{c}
\alpha_i \mid \beta, z, w, \sigma _{\alpha }^{2}, \zeta   \sim N\left( \widetilde{a}, \widetilde{A}\right), \quad \mathrm{for} \;\; i=1,\cdots,n, \vspace{2px}\\
\text{where} \;\; \widetilde{A}^{-1} = \left( \iota^{\prime}_{T_{i}} D_{\tau \sqrt{w_{i}}}^{-2} \iota _{T_{i}} +  \sigma_{\alpha}^{-2}  \right), \;\; \text{and} \;\; \widetilde{a}= \widetilde{A} \left(  \iota^{\prime} _{T_{i}} D_{\tau \sqrt{w_{i}}}^{-2}
    \left( z_{i}-X_{i}\beta - w_{i}    \theta  \right)  +   \sigma _{\alpha }^{-2}  \overline{m}_{i}^{\prime }\zeta  \right).
\end{array}
\label{eq11}
\end{equation}
%---------------------------------
\begin{equation}
\begin{array}{c}
w_{it} \mid \beta, \alpha_i , z_{it}   \sim GIG\left( \frac{1}{2}, \widetilde{\lambda}_{it}, \widetilde{\eta} \right), \quad \text{for} \; i=1,\cdots,n, \;\; \text{and} \;\; t=1,\cdots,T_i, \vspace{2px} \\
\text{where} \;\; \widetilde{\lambda}_{it} = \left(  \frac{   z_{it}-x^{\prime}_{it} \beta  -\alpha_i}{ \tau } \right)^2, \;\; \text{and} \;\; \widetilde{\eta}= \left( \frac{\theta^2}{\tau^2} + 2 \right).
\end{array}
\label{eq12}
\end{equation}
%---------------------------------
\begin{equation}
\begin{array}{c}
\sigma_{\alpha }^{2} \mid \alpha, \zeta   \sim IG\left( \frac{ \widetilde{c}_1}{2}, \frac{ \widetilde{d}_1}{2} \right), \vspace{2px} \\
\text{where} \;\; \widetilde{c}_{1} = (n + c_1), \;\; \text{and} \;\; \widetilde{d}_{1} = d_1 + \tsum\limits_{i=1}^{n}  \left( \alpha _{i}-\overline{m}_{i}^{\prime }\zeta \right)^{\prime }\left( \alpha_{i}-\overline{m}_{i}^{\prime }\zeta \right).
\end{array}
\label{eq13}
\end{equation}
%---------------------------------
\begin{equation}
\begin{array}{c}
\zeta \mid \alpha, \sigma_{\alpha }^{2} \sim N_{k-1}\left( \widetilde{\zeta}, \widetilde{\Sigma}_{\zeta}\right),\\
\text{where} \;\; \widetilde{\Sigma}^{-1}_{\zeta}  = \left( \sigma_{\alpha }^{-2}  \tsum\limits_{i=1}^{n} \overline{m}_{i} \overline{m}_{i}^{\prime } +  C^{-1}_{0} \right), \;\;
\text{and} \widetilde{\zeta} = \widetilde{\Sigma}_{\zeta} \left(\sigma_{\alpha }^{-2}  \tsum\limits_{i=1}^{n} \overline{m}_{i} \alpha^{\prime}_{i} +  C^{-1}_{0} \zeta_{0}\right).
\end{array}
\label{eq14}
\end{equation}
%---------------------------------

\newpage


\bibliographystyle{spbasic}
\bibliography{Manuscript}